\documentclass[aps,prd,onecolumn,nofootinbib,preprintnumbers,superscriptaddress,longbibliography,nobibnotes]{revtex4-2}

\usepackage{style}

\begin{document}

\preprint{FERMILAB-PUB-23-668-T, RIKEN-iTHEMS-Report-23}

\title{Searching for High Frequency Gravitational Waves with Phonons}

\author{Yonatan Kahn\,\orcidlink{0000-0002-9379-1838}}
\email{yfkahn@illinois.edu}
\affiliation{Illinois Center for Advanced Studies of the Universe and Department of Physics, University of Illinois Urbana-Champaign, Urbana, IL 61801, USA}

\author{Jan Sch\"utte-Engel\,\orcidlink{0000-0001-7905-237X}}
\email{janschue@berkeley.edu}
\affiliation{Department of Physics, University of California, Berkeley, CA 94720, USA}
\affiliation{RIKEN iTHEMS, Wako, Saitama 351-0198, Japan}
\affiliation{Illinois Center for Advanced Studies of the Universe and Department of Physics, University of Illinois Urbana-Champaign, Urbana, IL 61801, USA}

\author{Tanner Trickle\,\orcidlink{0000-0003-1371-4988}}
\email{ttrickle@fnal.gov}
\affiliation{Theoretical Physics Division, Fermi National Accelerator Laboratory, Batavia, IL 60510, USA}

\date{\today}

\begin{abstract}
    The gravitational wave (GW) spectrum at frequencies above a kHz is a largely unexplored frontier. We show that detectors with sensitivity to single-phonon excitations in crystal targets can search for GWs with frequencies, $\text{THz} \lesssim f \lesssim 100 \, \text{THz}$, corresponding to the range of optical phonon energies, $\text{meV} \lesssim \omega \lesssim 100 \, \text{meV}$. Such detectors are already being built to search for light dark matter (DM), and therefore sensitivity to high-frequency GWs will be achieved as a byproduct. We begin by deriving the absorption rate of a general GW signal into single phonons. We then focus on carefully defining the detector sensitivity to monochromatic and chirp signals, and compute the detector sensitivity for many proposed light DM detection targets. The detector sensitivity is then compared to the signal strength of candidate high-frequency GW sources, e.g., superradiant annihilation and black hole inspiral, as well as other recent detector proposals in the $\text{MHz} \lesssim f \lesssim 100 \, \text{THz}$ frequency range. With a judicious choice of target materials, a collection of detectors could optimistically achieve sensitivities to monochromatic signals with $h_0 \sim 10^{-23} - 10^{-25}$ over $\text{THz} \lesssim f \lesssim 100 \, \text{THz}$.
\end{abstract}

\maketitle
\newpage
\tableofcontents
\newpage

\section{Introduction}
\label{sec:intro}

The prediction of gravitational waves (GWs) by Einstein in 1916 sparked a century long search for their existence. The first indirect evidence came from measurements of the orbital decay of the Hulse-Taylor pulsar~\cite{Hulse:1974eb}, which were found consistent with GW emission~\cite{Taylor:1989sw}. Direct evidence of the existence of GWs would follow in 2016 when GWs from an inspiraling pair of black holes (BHs) were measured~\cite{LIGOScientific:2016aoc}. Even more recently the NANOGrav collaboration reported evidence of a stochastic GW background~\cite{NANOGrav:2023gor,NANOGrav:2023hvm,NANOGrav:2023hfp}. These initial direct detections mark the beginning of ``GW astronomy" as a viable method to study the universe. 

Today the search for GWs is rapidly expanding. Continued development of the aLIGO~\cite{Harry:2010zz, LIGOScientific:2014pky}, aVirgo~\cite{VIRGO:2014yos}, and KAGRA~\cite{Somiya:2011np, Aso:2013eba, KAGRA:2018plz, Michimura:2019cvl,KAGRA:2019htd} interferometers have improved the sensitivity to GWs in the $10 \, \text{Hz} \lesssim f \lesssim 10^3 \, \text{Hz}$ frequency range, and future ground-based interferometers, e.g., Cosmic Explorer~\cite{LIGOScientific:2016wof,Reitze:2019iox} and the Einstein Telescope~\cite{Punturo:2010zz,Hild:2010id,Sathyaprakash:2012jk,Maggiore:2019uih} will further enhance sensitivity in this frequency band. Future space-based interferometers, e.g., BBO~\cite{Corbin:2005ny, Crowder:2005nr, Harry:2006fi}, DECIGO~\cite{Seto:2001qf, Kawamura:2006up, Yagi:2011wg}, and LISA~\cite{LISA:2017pwj, Baker:2019nia}, will explore lower frequencies, $10^{-4} \, \text{Hz} \lesssim f \lesssim 1\, \text{Hz}$, with much longer arm lengths, while the gap between ground-based and space-based inteferometers could be covered by atom interferometers~\cite{Badurina:2019hst,MAGIS-100:2021etm}. Meanwhile, pulsar timing arrays (PTAs) such as the EPTA~\cite{Kramer:2013kea, Babak:2015lua, Lentati:2015qwp}, NANOGrav~\cite{McLaughlin:2013ira, NANOGRAV:2018hou,Aggarwal:2018mgp,Brazier:2019mmu}, PPTA~\cite{Shannon:2015ect, Manchester:2012za}, and the collective IPTA~\cite{Hobbs:2009yy,Manchester:2013ndt,Verbiest:2016vem, Hazboun:2018wpv} have been precisely monitoring many pulsars for $\mathcal{O}(10) \, \text{years}$, earning them sensitivity in the $10^{-9} \, \text{Hz} \lesssim f \lesssim 10^{-6} \, \text{Hz}$ frequency range. While this leaves the ``$\mu$Hz gap" in the $10^{-6} \, \text{Hz} \lesssim f \lesssim 10^{-4} \, \text{Hz}$ frequency range, there are proposals to bridge this gap, e.g., using asteroids in the solar system~\cite{Fedderke:2021kuy} or precisely ranging the moon or orbiting satellites~\cite{Blas:2021mqw}. Clearly, the future of GW detection at frequencies $f \lesssim 10^3 \, \text{Hz}$ is promising.

At high frequencies above the ground based interferometer band, many different proposals have been put forward; see Refs.~\cite{Aggarwal:2020olq,Domcke:2023qle} for some recent reviews. Optically levitated sensors~\cite{Aggarwal:2020umq} and experiments utilizing the M\"{o}ssbaur effect~\cite{Gao:2023ggo} may be sensitive in the $10^3 \, \text{Hz} \lesssim f \lesssim 10^6 \, \text{Hz}$ range, bulk acoustic wave resonators~\cite{Goryachev:2014yra, Goryachev:2014nna, Goryachev:2021zzn} may be sensitive in the $10^6 \, \text{Hz} \lesssim f \lesssim 10^9 \, \text{Hz}$ range (with demonstrated sensitivity at $f \sim 5 \times 10^6 \, \text{Hz}$~\cite{Goryachev:2021zzn}), collections of different short arm laser interferometric experiments~\cite{Nishizawa:2007tn, Akutsu:2008qv, Holometer:2016qoh, Bailes:2019oma, Ackley:2020atn} can also explore the $10^3 \, \text{Hz} \lesssim f \lesssim 10^9 \, \text{Hz}$ range, and graviton to magnon conversion has been shown to have potential for searching at $f \sim 10^{10} \, \text{Hz}$~\cite{Ito:2019wcb}. Proposals such as MAGO 2.0~\cite{Berlin:2023grv} and using optical clocks~\cite{Bringmann:2023gba} may be sensitive to the $10^3 \, \text{Hz} \lesssim f \lesssim 10^9 \, \text{Hz}$ range with single experiments. Furthermore, it has been shown that existing experiments searching for axion dark matter (DM) with, e.g., microwave cavities~\cite{Berlin:2021txa} and LC circuits~\cite{Domcke:2023bat}, are also sensitive to high-frequency GWs, allowing for synergistic searches for DM and GWs with the same detector in the $10^3 \, \text{Hz} \lesssim f \lesssim 10^{10} \, \text{Hz}$ range. For frequencies above $10^{10} \, \text{Hz}$, and below $10^{14} \, \text{Hz}$ where light shining through wall experiments, e.g., ALPS,~\cite{Ehret:2010mh}, CAST~\cite{CAST:2017uph}, and OSQAR~\cite{OSQAR:2015qdv} have sensitivity~\cite{Ejlli:2019bqj}, there is an absence of searches. 

In this work we show that high-frequency GWs can be detected via conversion into single phonons in crystal targets. Phonons are quanta of lattice vibrations, and most crystal targets have gapped phonon modes, with energies in the $1 \, \text{meV} \lesssim \omega \lesssim 100 \, \text{meV}$ range, corresponding to frequencies $10^{12} \, \text{Hz} \lesssim f \lesssim 10^{14} \, \text{Hz}$. Because these modes are gapped, an incoming GW can be kinematically matched to the phonon dispersion relation, and therefore can resonantly vibrate the ions in the lattice, creating a phonon. This is somewhat analogous to an incoming photon converting to a phonon in a ``polar" crystal, i.e., crystals with charged ions in the unit cell. If an incoming photon is kinematically matched to a gapped phonon mode in a polar target, the photon will resonantly drive dipole oscillations, which is equivalent to exciting the gapped phonon mode.\footnote{Gapped phonon modes exist in any crystal where the number of atoms in the unit cell is greater than one. They are often referred to as ``optical" phonons, since they couple to light in polar targets. However they exist in non-polar, e.g., Silicon and Germanium, targets well.}

The scarcity of high-frequency GW detectors is in part due to the difficulty of generating sufficiently strong GWs with astrophysical or cosmological sources. The generic problem of creating abundant high-frequency GWs is that the rapid oscillations necessary to generate high frequencies can only be achieved with astrophysically small masses, thereby limiting the signal strength. For example, this is precisely the difficulty in generating high-frequency GWs from the inspiral of two BHs. As we discuss in Sec.~\ref{subsec:mono_sources}, the maximum BH mass, $M_\text{max}$, which can generate GWs at a frequency $\omega$ is $M_\text{max} \sim 10^{-8} M_\odot \left( \text{THz} / \omega \right)$. This is a much smaller mass, corresponding to a much smaller signal, than the $\mathcal{O}(10) \, M_\odot$ BHs LIGO is sensitive to.

However, it is not physically impossible to generate detectable high-frequency GWs, and studying the sensitivity of current technology, likely built for other reasons, e.g., DM direct detection, provides excellent motivation for further studies of high-frequency GW sources. Indeed, perhaps the most important feature of our proposed detection scheme is that the detectors are \emph{already} being built for DM direct detection experiments. Direct detection of DM by single-phonon excitations has been shown to be a promising route to search for both scattering of low-mass (sub-MeV) DM~\cite{Schutz:2016tid,Knapen:2017ekk,Griffin:2018bjn,Trickle:2019nya,Cox:2019cod,Kurinsky:2019pgb,Griffin:2019mvc,Trickle:2020oki,Griffin:2020lgd,Coskuner:2021qxo,Knapen:2021bwg,Taufertshofer:2023rgq,Romao:2023zqf} and absorption of sub-eV bosonic DM~\cite{Knapen:2017ekk,Griffin:2018bjn,Mitridate:2020kly,Knapen:2021bwg,Berlin:2023ppd,Mitridate:2023izi}. The TESSARACT experiment~\cite{Chang2020} plans to utilize both \ce{Al2O3} and \ce{GaAs} as targets. Therefore, similar to the repurposing of axion DM detectors as GW detectors~\cite{Berlin:2021txa, Domcke:2023qle}, any future DM direct detection experiment utilizing single-phonon excitations can directly be used as a GW detector.

This paper is organized as follows. In Sec.~\ref{sec:theory_derivation} we derive the GW absorption rate into single phonons in two steps. First, in Sec.~\ref{subsec:GR_force_derivation}, we summarize the general-relativistic derivation for the forces on a lattice of point masses due to an incoming GW. Then in Sec.~\ref{subsec:QM_phonon_abs_rate}, using the interaction Hamiltonian generated by the GW force, we derive the absorption rate into single-phonon excitations in both polar and non-polar targets. In Sec.~\ref{sec:results} we begin by carefully defining the detector sensitivity to \textit{deterministic} signals, i.e., those for which the GW strain is non-stochastic. We then discuss the detector sensitivity for a wide variety of target materials previously considered in the context of DM direct detection~\cite{Griffin:2019mvc}. In Sec.~\ref{subsec:mono_sources} we discuss the general difficulty with generating high-frequency GWs, and then consider specific examples of potential high-frequency sources, such as superradiant annihilation~\ref{subsubsec:superradiance} and BH inspiral~\ref{subsubsec:BH_merger}, carefully understanding their signal strength in the context of counting experiments. We conclude in Sec.~\ref{sec:conclusion}. A brief discussion of constraints on stochastic sources comprises Appendix~\ref{app:sto_constraints}.

\section{Theoretical Foundation}
\label{sec:theory_derivation}

The interaction of GWs with solid objects can be described in multiple ways. If the wavelength of the incoming GW is much larger than the detector, the deformation of the detector can be described using the theory of elasticity~\cite{Weber:1960zz,Landau:1986aog,maggiore}. However, when the frequency of the GW is much larger, comparable to the $\mathcal{O}(\text{meV}-100\, \text{meV})$ gapped phonons, the low energy effective theory of elasticity is no longer appropriate.
The GW wavelengths, $\mathcal{O}(10\,\mu \text{m} - \text{mm})$ are smaller than the size of bulk single crystals. While much smaller than the size of the detecting crystal, these wavelengths are still much larger than the interatomic spacing, $\mathcal{O}(10^{-10}\,\text{m})$. Therefore, similar to how $\mathcal{O}(\text{meV})$ photons couple to the dipole moment of a unit cell in polar targets to generate phonons, the effect of an incoming GW can be intuitively understood as a coupling to the quadrupole mass moment of the unit cell, which then generates phonons. Since the GW interaction is more similar to the photon interaction, computing the number of phonons produced from GWs proceeds more naturally by inheriting methods from particle and condensed matter physics.

The purpose of this section is to derive the absorption rate of incoming GWs into single-phonon excitations. We begin in Sec.~\ref{subsec:GR_force_derivation} with a review of the derivation of the force on a point mass within general relativity, and then generalize to a lattice of point masses. Furthermore we detail the necessary assumptions needed to treat the effect of the incoming GW on each point mass as a classical force. The potential energy associated with this force is the starting point of Sec.~\ref{subsec:QM_phonon_abs_rate}, and defines the interaction Hamiltonian. For non-polar targets, the focus of Sec.~\ref{subsubsec:non_polar_targets}, the absorption rate then follows simply from Fermi's Golden Rule, and we derive the necessary matrix elements. In polar targets, Sec.~\ref{subsubsec:polar_targets}, the phonons mix with the photon, complicating the absorption rate derivation. To compute the absorption rate for these targets we utilize the formalism from Refs.~\cite{Hardy:2016kme,Mitridate:2021ctr,Mitridate:2023izi}, which has carefully accounted for these mixing effects in the context of light DM absorption. We will work in natural units, $\hbar = c = 1$, and use a mostly positive, $(-,+,+,+)$, metric signature.

\subsection{General Relativistic Forces}
\label{subsec:GR_force_derivation}

Our goal is to understand how a weak GW interacts with a crystal lattice, and therefore the natural starting point is understanding how a single mass interacts with a GW. While this is a textbook discussion~\cite{Misner:1973prb,maggiore}, the general coordinate invariance of GR renders this a delicate subject. For example, in transverse-traceless (TT) coordinates the coordinate position of a single mass is unaffected by a passing GW. However this is simply an artifact of TT coordinates, reminding us that coordinates have no inherent meaning in GR; physics is encoded in coordinate invariants, e.g., proper distances. Fermi-normal (FN) coordinates equip the coordinates with physical meaning by defining the coordinates as the proper distance to an observer.\footnote{Recent literature~\cite{Rakhmanov:2014noa,Berlin:2021txa} has explicitly demonstrated the applicability of FN coordinates beyond the usual long wavelength approximation, $\lambda \gg L$, where $\lambda$ is the GW wavelength, and $L$ is the size of the detector, to $\lambda \gg \sqrt{h} L$ which is easily satisfied for the detectors considered here.}

Following Ref.~\cite{Rakhmanov:2014noa}, the equation of motion for a test mass in FN coordinates, assuming a non-relativistic mass and observer at the origin, is given by,
\begin{align}
    \frac{d^2 x^i}{d t^2} \approx \frac{1}{2} \, \ddot{h}_{ij} \, x^j  \, ,
    \label{eq:EOM_approx}
\end{align}
to linear order in $h$, where the GW perturbation, $h_{\mu \nu}$, is defined as a perturbative addition to the metric, $g_{\mu \nu} \approx \eta_{\mu \nu} + h_{\mu \nu}$, and $\eta_{\mu \nu}$ is the Minkowski metric. Note that $h_{ij}$ in defined in TT coordinates and the right hand side of Eq.~\eqref{eq:EOM_approx} is coordinate independent to linear order in $h$: TT and FN coordinates are related by $\mathcal{O}(h)$~\cite{Rakhmanov:2014noa}, and therefore a coordinate transformation would transform the right hand side of Eq.~\eqref{eq:EOM_approx} from $\mathcal{O}(h) \rightarrow \mathcal{O}(h) + \mathcal{O}(h^2)$. From Eq.~\eqref{eq:EOM_approx} we see that a passing GW acts like a fictitious tidal force, $F^i$,
\begin{align}
    F^i = \frac{m}{2} \, \ddot{h}_{ij} \, x^j \, .
    \label{eq:final_GR_force}
\end{align}
This result is straightforwardly generalized to a crystal lattice, i.e., each mass experiences a force, $F_I^i = m_I \, \ddot{h}_{ij} \, x^j_I /2$, where $I$ indexes the mass.

In addition to gravitational forces, the crystal lattice dynamics are of course influenced by the electromagnetic forces binding the crystal. These determine the leading order, $\mathcal{O}(h^0)$, contributions to the position and are simply the equilibrium positions of the ions (or, equivalently, atoms in non-polar targets), $x^i_I \approx x_{I, 0}^i$. Perturbations around this equilibrium, $u_I^i(t)$, induced by the GW are then driven by the force in Eq.~\eqref{eq:final_GR_force}, and back to equilibrium by the crystal harmonic ``spring" forces. The equation of motion, at $\mathcal{O}(h)$, is then given by,
\begin{align}
    m_I \frac{d^2 u^i_I}{d t^2} + \sum_J V_{IJ}^{ij} \, u_J^j =  \frac{m_I}{2} \, \ddot{h}_{ij} \, x_{I, 0}^j \, ,
    \label{eq:EOM_non_grav_2}
\end{align}
where $V_{IJ}$ is the spring constant connecting mass $I$ to mass $J$. Note that the $V_{IJ}^{ij}$ only need to be determined to $\mathcal{O}(h^0)$. The perturbations themselves are $\mathcal{O}(h)$, and therefore, analogous to the driving GW force, a coordinate transformation would only generate shifts from $\mathcal{O}(h)$ to $\mathcal{O}(h) + \mathcal{O}(h^2)$. Therefore the effect of a GW on a lattice of masses is to introduce a fictitious force, Eq.~\eqref{eq:final_GR_force}, which can drive the phonon modes in a crystal, analogous to a passing electromagnetic wave. 

\subsection{Absorption Rate Calculation}
\label{subsec:QM_phonon_abs_rate}

Since phonons are the quanta of lattice vibrations, the natural formalism to discuss single-phonon absorption rates is quantum mechanics. In calculating the absorption rate, the incoming GW may be treated as either a classical GW background or an incoming graviton~\cite{Carney:2023nzz}, analogous to equivalence of describing a photon classically or quantum mechanically when considering light-matter interactions. We will choose the latter option for consistency with the phonon description. Lattice vibrations are described as oscillations around an equilibrium position; the position of the $j^\text{th}$ ion in the $\ell^\text{th}$ unit cell of the crystal is
\begin{align}
    \mathbf{x}_{\ell j}(t) = \mathbf{x}_{\ell j}^0 + \mathbf{u}_{\ell j}(t) = \mathbf{r}_{\ell}^0 + \mathbf{x}^0_j + \mathbf{u}_{\ell j}(t) \, , \label{eq:position_equation}
\end{align}
where $\mathbf{x}_{\ell j}^0$ is the equilibrium position, $\mathbf{r}^0_\ell$ is the lattice vector of the $\ell^\text{th}$ unit cell, $\mathbf{x}^0_j$ is the equilibrium position relative to the center of the unit cell, and $\mathbf{u}_{\ell j}(t)$ is the displacement of the ion away from equilibrium. When quantizing the lattice vibrations~\cite{Wallace_1972}, $\mathbf{u}_{\ell j}(t)$ can be expanded in terms of phonon raising and lowering operators:
\begin{align}
    \mathbf{u}_{\ell j}(t) & = \frac{1}{\sqrt{2 N m_j}} \sum_{\nu \mathbf{k}} e^{-i \omega_{\nu \mathbf{k}} t} \, \frac{e^{i \mathbf{k} \cdot \mathbf{x}^0_{\ell j}}}{\sqrt{\omega_{\nu \mathbf{k}}}} \left( a_{\nu \mathbf{k}} + a_{\nu, -\mathbf{k}}^\dagger \right) \bm{\epsilon}_{j \nu\mathbf{k}} \, , \label{eq:phonon_operator}
\end{align}
where $N$ is the number of unit cells in the lattice, $m_j$ is the mass of the $j^\text{th}$ ion, $\nu$ indexes the band number, $\mathbf{k}$ indexes the first Brillouin zone (1BZ) momentum vectors, $\omega_{\nu \mathbf{k}}$ is the phonon energy, $a_{\nu\mathbf{k}}$ and $a^\dagger_{\nu\mathbf{k}}$ are the raising and lowering operators, respectively, which satisfy the canonical commutation relations, $[ a_{\nu\mathbf{k}}, a^\dagger_{\nu'\mathbf{k}'}] = \delta_{\nu \nu'} \delta_{\mathbf{k} \mathbf{k}'}$; and $\bm{\epsilon}_{\nu j \mathbf{k}}$ are the phonon polarization vectors and satisfy $\bm{\epsilon}_{\nu j -\mathbf{k}}=\bm{\epsilon}^*_{\nu j \mathbf{k}}$.

The interaction Hamiltonian between phonons and a GW is determined by the coupling of $\mathbf{u}_{\ell j}$ to $h_{ij}$, which is due to the force in Eq.~\eqref{eq:EOM_non_grav_2},
\begin{align}
    \delta H = -\frac{1}{2} \sum_{\ell j} m_j \, u^{i}_{\ell j} \, \ddot{h}_{ik} \,  x^{0, k}_{\ell j} \, . \label{eq:GW_phonon_interaction_Hamiltonian_0}
\end{align}
This interaction Hamiltonian is the basis for the absorption rate calculation. At first sight one might simply wish to apply Fermi's Golden Rule to compute the graviton induced single-phonon transition rate. However, in \textit{polar} targets, where the ions have charge, $Q_j$, the phonon mixes with the photon, and screening can occur. Therefore we split the absorption rate in to two calculations. In Sec.~\ref{subsubsec:non_polar_targets} we consider \textit{non-polar} materials, where $Q_j = 0$, and Fermi's Golden Rule may be straightforwardly applied. In Sec.~\ref{subsubsec:polar_targets} we consider general materials, and compute the absorption rate using the optical theorem, following the derivation in Ref.~\cite{Mitridate:2023izi}. While the approach in Sec.~\ref{subsubsec:polar_targets} is more general, and reproduces the results in Sec.~\ref{subsubsec:non_polar_targets} in the limit of a non-polar material, it is also more technically involved.

Before continuing it is worthwhile to introduce notation common to both sections. It will be useful to work with the canonically-normalized GW field as opposed to the perturbations in the metric directly. This involves a rescaling, $h_{ij} \rightarrow \sqrt{16 \pi G} \, h_{ij}$,  
which transforms the interaction Hamiltonian in Eq.~\eqref{eq:GW_phonon_interaction_Hamiltonian_0} to
\begin{align}
    \delta H =- \sqrt{4 \pi G} \sum_{\ell j} m_{j} \, u^{i}_{\ell j} \, \ddot{h}_{ik} \,  x^{0, k}_{\ell j} \, , \label{eq:GW_phonon_interaction_Hamiltonian}
\end{align}
where $h_{ij}$ now has mass dimension one, and is quantized in the TT frame as,
\begin{align}
    h^{ij}(\mathbf{x}, t) & = \frac{1}{\sqrt{V}} \sum_{\lambda \mathbf{p}} e^{- i \omega_{\mathbf{p}} t} \frac{e^{ i \mathbf{p} \cdot \mathbf{x}}}{\sqrt{\omega_\mathbf{p}}} \left( a_{\lambda\mathbf{p}} + a^\dagger_{\lambda, - \mathbf{p}}\right) e^{ij}_\lambda(\hat{\mathbf{p}}) \, , \label{eq:h_quantized}
\end{align}
where $\omega_{\mathbf{p}} = |\mathbf{p}|$, and the overall normalization is found by equating the energy density, $\rho_\text{GW} = \langle \dot{h}_{ij} \dot{h}^{ij} \rangle / 2$, to the quantum mechanical Hamiltonian density, $H / V = (1/V) \sum_{\lambda\mathbf{p}} \omega_{\mathbf{p}} \, a_{\lambda\mathbf{p}}^\dagger a_{\lambda\mathbf{p}}$~\cite{Boughn:2006st} with polarization vectors normalized to $e_{ij}^\lambda \, e^{ij}_{\lambda'} = \delta^{\lambda \lambda'}$.\footnote{This choice of normalization for the GW polarization tensors is different than, e.g., Ref.~\cite{maggiore}, which chooses to normalize as $e_{ij}^\lambda \, e^{ij}_{\lambda'} = 2 \delta^{\lambda \lambda'}$.} Additionally we take the incoming GW to have four momentum, $Q^\mu = (\omega, \mathbf{q}) = (\omega, \omega \hat{\mathbf{n}})$, where $\omega = 2 \pi f$ is the GW frequency and $\mathbf{q} = \omega \hat{\mathbf{n}}$ is the GW momentum.

The absorption rate per incoming graviton of polarization $\lambda$ is given by $\Gamma_\lambda$. The total absorption rate, per detector exposure, for any kind of GW with polarization $\lambda$, is given by, 
\begin{align}
    R_\lambda = \frac{1}{\rho_T} \int \frac{d n^\lambda_\text{GW}}{d f d \hat{\mathbf{n}}} \, \Gamma_\lambda(f, \hat{\mathbf{n}}) \, d f d\hat{\mathbf{n}} \, ,
    \label{eq:rate_1}
\end{align}
where $\rho_T$ is the target mass density, and $n^\lambda_\text{GW}$ is the number density of the $\lambda^\text{th}$ polarization of GWs. The differential energy density is related to the number density by $d \rho^\lambda_\text{GW} = 2 \pi f d n_\text{GW}^\lambda$. Assuming the incoming GW is isotropic and independent of polarization,
\begin{align}
    \frac{d n^\lambda_\text{GW}}{ d f d\hat{\mathbf{n}}} = \frac{1}{2 \pi f} \frac{d \rho_\text{GW}^\lambda}{d f d \hat{\mathbf{n}}} = \frac{1}{2 \pi f} \frac{1}{4\pi} \frac{d \rho_\text{GW}^\lambda}{d f } = \frac{1}{2 \pi f} \frac{1}{4 \pi} \frac{1}{2} \frac{d \rho_\text{GW}}{d f} = \frac{\rho_c}{16 \pi^2 f^2} \Omega_\text{GW}(f) \, ,
    \label{eq:isotropic_independent_GW_approximation}
\end{align}
where $\rho_c = 3 H_0^2 / 8 \pi G$ is the critical density, $H_0$ is the Hubble constant today and $ \Omega_\text{GW}(f) \equiv (1/\rho_c)\, d \rho_\text{GW}/d \log f$.
Substituting Eq.~\eqref{eq:isotropic_independent_GW_approximation} in to Eq.~\eqref{eq:rate_1}, and averaging over the GW polarizations to find the total, averaged, GW absorption rate, $R$, gives
\begin{align}
    R & \equiv \frac{1}{2} \sum_\lambda R_\lambda = \frac{1}{4 \pi} \frac{\rho_c}{\rho_T} \int  \frac{1}{f^2} \, \Omega_\text{GW}(f) \, \Gamma(f) \, df \label{eq:total_averaged_GW_abs_rate} \\
    \Gamma(f) & \equiv \frac{1}{8 \pi} \sum_\lambda \int \Gamma_\lambda(f, \hat{\mathbf{n}}) \, d\hat{\mathbf{n}} \, ,
\end{align}
where we have defined $\Gamma(f)$ as the polarization and angularly averaged absorption rate per GW. Therefore to compute $R$ in Eq.~\eqref{eq:total_averaged_GW_abs_rate}, we need to compute $\Gamma_\lambda$, which will be the focus of Secs.~\ref{subsubsec:non_polar_targets} and~\ref{subsubsec:polar_targets}. 

For both polar and non-polar targets, in order to compute $\Gamma_\lambda$ the phonon energies $\omega_{\nu\mathbf{k}}$, polarization eigenvectors $\bm{\epsilon}_{\nu j \mathbf{k}}$, and equilibrium ion positions $\mathbf{x}^0_j$ are needed. We calculate these with first-principles methods similar to Refs.~\cite{Trickle:2019nya,Griffin:2019mvc,Trickle:2020oki,Mitridate:2020kly,Coskuner:2021qxo,Mitridate:2023izi}, which we briefly summarize here. Using \textsf{VASP}~\cite{Kresse_1993,Kresse_1994,Kresse_1996} the ions and electrons in the lattice are relaxed to their minimum energy configuration, which sets the equilibrium positions, and perturbing the ions away from these minima allows calculation of the ``force constants", $\mathbf{V}_{\ell \ell' j j'}$, which define the harmonic phonon Hamiltonian,
\begin{align}
    H_\text{ph} & = \frac{1}{2} \sum_{\ell j}  m_j \, \dot{\mathbf{u}}_{\ell j} \cdot \dot{\mathbf{u}}_{\ell j} + \frac{1}{2}\sum_{\ell \ell' j j'} \mathbf{u}_{\ell' j'} \cdot \mathbf{V}_{\ell \ell' j j'} \cdot \mathbf{u}_{\ell j} \, .
    \label{eq:phonon_hamiltonian}
\end{align}
Then $\mathbf{V}_{\ell \ell' j j'}$ is diagonalized with the help of \textsf{phonopy}~\cite{phonopy-phono3py-JPSJ,phonopy-phono3py-JPCM} to find the phonon eigensystem, and the final absorption rate, $R$ in Eq.~\eqref{eq:total_averaged_GW_abs_rate}, is computed with the help of \textsf{PhonoDark-abs}~\cite{Mitridate:2023izi}.

\subsubsection{Non-Polar Targets}
\label{subsubsec:non_polar_targets}

Since the atoms at each site in non-polar targets are electrically neutral, there is no phonon-photon mixing and the absorption rate can be computed straightforwardly with Fermi's Golden Rule. Specifically, we will compute the transition rate due to the interaction Hamiltonian $\delta H$ in Eq.~\eqref{eq:GW_phonon_interaction_Hamiltonian} from an initial state, $|I \rangle = | \lambda, f, \hat{\mathbf{n}} \rangle \otimes | 0 \rangle$, containing zero phonons and an incoming graviton, to a final state, $| F \rangle = | 0 \rangle \otimes | \nu, \mathbf{k} \rangle$, containing zero gravitons and one phonon indexed by its band number $\nu$ and crystal momentum vector $\mathbf{k}$.  This rate is given by
\begin{align}
    \Gamma_\lambda(f, \hat{\mathbf{n}}) =  2 \pi \sum_{\nu \mathbf{k}} |\langle F | \delta H | I \rangle |^2  \delta(\omega - \omega_{\nu \mathbf{k}}) \, .
    \label{eq:fermi_golden_rule}
\end{align}
where $\delta H$ is evaluated at $t = 0$, since the time dependence has been factored out to provide the energy conserving delta function as in ordinary time-dependent perturbation theory, and $\langle I | I \rangle = \langle F | F \rangle = 1$. The phonon contribution to the matrix element is computed using Eq.~\eqref{eq:phonon_operator},
\begin{align}
    \langle \nu, \mathbf{k} | \mathbf{u}_{\ell j} | 0 \rangle & = \frac{e^{- i \mathbf{k} \cdot \mathbf{x}^0_{\ell j}}}{\sqrt{2 N m_j \omega_{\nu \mathbf{k}}}} \bm{\epsilon}^*_{j \nu \mathbf{k}} \equiv \frac{e^{-i \mathbf{k} \cdot \mathbf{x}^0_{\ell j}}}{\sqrt{N}} \mathbf{T}_{j \nu \mathbf{k}} \approx \frac{e^{-i \mathbf{k} \cdot \mathbf{r}^0_{\ell}}}{\sqrt{N}} \mathbf{T}_{j \nu \mathbf{k}} \, , \label{eq:phonon_matrix_element}
\end{align}
where $\mathbf{u}_{\ell j} \equiv \mathbf{u}_{\ell j}(t = 0)$, $| \nu, \mathbf{k} \rangle = a_{\nu\mathbf{k}}^\dagger| 0 \rangle$, we have defined the \textit{phonon transition matrix element}, $\mathbf{T}_{j\nu\mathbf{k}}$, and used the long-wavelength approximation $|\mathbf{k} \cdot \mathbf{x}_j^0| \ll 1$, which is true for the momentum transfers of interest here since $|\mathbf{k}| \sim \omega \sim \text{meV}$ and $|\mathbf{x}_j^0| \sim \text{\AA} \sim (\text{keV})^{-1}$.

The graviton contribution to $\langle F | \delta H | I \rangle$ is then given by evaluating the matrix element of $\ddot{h}_{ij}(\mathbf{x}^0_{\ell j}) \equiv \ddot{h}_{ij}(\mathbf{x}^0_{\ell j}, 0)$ in Eq.~\eqref{eq:h_quantized}, using the quantization of the graviton field from Eq.~\eqref{eq:h_quantized},
\begin{align}
    \langle 0 | \, \ddot{h}^{ij}(\mathbf{x}^0_{\ell j}) \, | \lambda, f, \hat{\mathbf{n}} \rangle & = - \sqrt{\frac{ \omega^3 }{V}} \, e^{i \mathbf{q} \cdot \mathbf{x}^0_{\ell j}} \, e_\lambda^{ij}(\hat{\mathbf{n}}) \approx - \sqrt{\frac{ \omega^3 }{V}} \, e^{i \mathbf{q} \cdot \mathbf{r}^0_{\ell}} \, e_\lambda^{ij}(\hat{\mathbf{n}}) \, , \label{eq:GW_matrix_element_1}
\end{align}
where $|\lambda, f, \hat{\mathbf{n}} \rangle = a_{\lambda\mathbf{q}}^\dagger |0 \rangle$ and in the last step we made the same long-wavelength approximation $|\mathbf{q} \cdot \mathbf{x}_j^0| \ll 1$ as in Eq.~\eqref{eq:phonon_matrix_element}.

Substituting Eqs.~\eqref{eq:phonon_matrix_element} and~\eqref{eq:GW_matrix_element_1} into $\langle F | \delta H | I  \rangle$ gives
\begin{align}
    \langle F | \delta H | I \rangle & = \frac{\sqrt{4 \pi G}}{N} \sqrt{\frac{\omega^3}{\Omega}} e_{ik}^\lambda(\hat{\mathbf{n}}) \sum_{\ell j}e^{i (\mathbf{q} - \mathbf{k}) \cdot \mathbf{r}^0_{\ell}} \, m_j \, x^{0, i}_{\ell j} \, T^k_{j\nu \mathbf{k}} \, ,
    \label{eq:matrix_element_1}
\end{align}
where $\Omega = V / N$ is the volume of the unit cell. Eq.~\eqref{eq:matrix_element_1} can be simplified further since the phonon transition matrix elements satisfy the ``coupling to mass"~\cite{Knapen:2017ekk,Griffin:2018bjn,Cox:2019cod,Mitridate:2020kly} sum rule, $\sum_j m_j \mathbf{T}_{j \nu \mathbf{k}} = 0$, when $\nu$ runs over the gapped, optical modes. Therefore, when expanding $\mathbf{x}^0_{\ell j}$ the term proportional to $\mathbf{r}_{\ell}^0$  vanishes, and using $\sum_{\ell} e^{i (\mathbf{q} - \mathbf{k}) \cdot \mathbf{r}_{\ell}} = N \delta_{\mathbf{q}, \mathbf{k}}$ further simplifies Eq.~\eqref{eq:matrix_element_1} to\footnote{Note that the coupling to mass effect also enforces independence of the final result from the absolute position of the unit cell center. To see this, imagine choosing a new center, shifted by some $\Delta \mathbf{x}$, such that $\mathbf{x}^{0}_j \rightarrow \mathbf{x}^0_j + \Delta \mathbf{x}$. The contribution to $\langle F | \delta H | I \rangle$ in Eq.~\eqref{eq:matrix_element_3} from $\Delta \mathbf{x}$ then vanishes due to the coupling to mass effect.}
\begin{align}
    \langle F | \delta H | I \rangle & = \delta_{\mathbf{q}, \mathbf{k}} \sqrt{\frac{4 \pi G \omega^3}{\Omega}} e_{ik}^\lambda(\hat{\mathbf{n}}) \sum_{j} m_j \, x^{0, i}_{j} \, T^k_{j \nu \mathbf{k}} \, .
    \label{eq:matrix_element_3}
\end{align}
Lastly, substituting Eq.~\eqref{eq:matrix_element_3} in to Eq.~\eqref{eq:fermi_golden_rule}, we obtain
\begin{align}
    \Gamma_\lambda(f, \hat{\mathbf{n}}) = \frac{8 \pi^2 G \omega^3}{\Omega} \sum_\nu \left( \sum_{j} m_j \, x^{0, i}_{j} \, e^\lambda_{ik} \, T^k_{j \nu \mathbf{q}} \right) \left( \sum_{j} m_j \, x^{0, i}_{j} \, e^\lambda_{ik} \, T^{k}_{j \nu \mathbf{q}} \right)^* \delta(\omega - \omega_{\nu}) \, , \label{eq:absorption_rate_per_graviton}
\end{align}
where we have approximated $\omega_{\nu\mathbf{q}} \approx \omega_{\nu}$, as appropriate for momentum transfers well within the 1BZ.

\subsubsection{Polar Targets}
\label{subsubsec:polar_targets}

If the ions on the lattice sites have net electric charge, then they will also couple to the photon in addition to any incoming GW. This mixing introduces screening effects which frequently arise when studying absorption or scattering of light DM~\cite{An:2014twa,Hardy:2016kme,Hochberg:2016ajh,Knapen:2017ekk,Mitridate:2021ctr,Mitridate:2023izi,Berlin:2023ppd}. Accounting for this screening will require introducing some formalism which has appeared in Refs.~\cite{Hardy:2016kme,Mitridate:2021ctr,Mitridate:2023izi}. While technically more complex, this formalism will also apply to non-polar targets and provide a rigorous method for smearing the delta function in Eq.~\eqref{eq:absorption_rate_per_graviton}, in addition to being able to account for mixing effects in polar targets. 

Consider an effective Lagrangian containing the graviton and photon fields, which are mixed by self-energies containing states in the medium, e.g., phonons or electrons,
\begin{align}
    \mathcal{L} & = \mathcal{L}^0_h + \mathcal{L}^0_A - \frac{1}{2}
    \begin{pmatrix} h_{\mu \nu} & A_\rho \end{pmatrix} 
        \begin{pmatrix} \Pi_{hh}^{\mu \nu, \, \mu' \nu'} & \Pi_{h A}^{\mu \nu,\, \rho'} \\ \Pi_{A h}^{\rho,\, \mu' \nu'} & \Pi_{AA}^{\rho\rho'} \end{pmatrix}
    \begin{pmatrix} h_{\mu' \nu'} \\ A_{\rho'} \end{pmatrix}  \label{eq:mixed_Lagrangian} \\ 
    \mathcal{L}_h^0 & = -\frac{1}{2} \left( \partial^\mu h^{\nu \rho} \partial_\mu h_{\nu \rho} - 2 \, \partial^\mu h^{\nu \rho} \partial_\nu h_{\mu \rho} + 2 \, \partial^\mu h \partial^\nu h_{\nu \mu} - \partial^\mu h \partial_\mu h \right) \\ 
    \mathcal{L}_A^0 & = - \frac{1}{4} F^{\mu \nu} F_{\mu \nu} \, ,
\end{align}
where $\mathcal{L}_h^0, \mathcal{L}_A^0$ are the free graviton and photon Lagrangians in vacuum, respectively, and the second term in Eq.~\eqref{eq:mixed_Lagrangian} contains the self-energies, $\Pi$, which are 1PI diagrams induced by the medium~\cite{Hardy:2016kme,Mitridate:2021ctr,Mitridate:2023izi}. For example, the single-phonon contribution to $\Pi_{hh}$ is given by the diagram,
\vspace{1em}
\begin{align}
    \begin{fmffile}{Pi_hh_ph}
    \begin{fmfgraph*}(100,20)
        \fmfleft{i} \fmfright{f}
        \fmf{dbl_wiggly,label=$\overset{Q}{\longrightarrow}$,label.dist=-0.175w}{i,m1}
        \fmf{double}{m1,m2}
        \fmf{dbl_wiggly}{m2,f}
        \fmfblob{.05w}{m1,m2}
        \fmflabel{$\Pi_{hh}(Q) \, : \quad h$}{i}
        \fmflabel{$h$}{f}
    \end{fmfgraph*}
    \end{fmffile} \label{eq:example_feynman_diagram}
\end{align}
where the internal double solid line indicates a phonon propagator, the vertex Feynman rule is determined by the interaction Hamiltonian in Eq.~\eqref{eq:GW_phonon_interaction_Hamiltonian}, and we have left indicies implicit for simplicity. The single-phonon contribution to the other self-energies, e.g., $\Pi_{hA}, \Pi_{AA}$, are analogous diagrams with the corresponding external $h$ replaced with $A$.

The in-medium states are those which diagonalize the Lagrangian in Eq.~\eqref{eq:mixed_Lagrangian}, and can be separated into unmixed ``graviton-like" ($\hat{h}$) and ``photon-like" ($\hat{A}$) states, since the couplings between the graviton and photon are perturbatively induced via the medium. The absorption rate of an incoming vacuum graviton is then, approximately, given by the absorption rate of its graviton-like counterpart, which can be computed with the optical theorem,
\begin{align}
    \Gamma_\lambda(f, \hat{\mathbf{n}}) \approx \Gamma_{\hat{h}}^\lambda(f, \hat{\mathbf{n}}) & = - \frac{1}{\omega} \text{Im} \left[ \Pi_{\hat{h}\hat{h}}^\lambda(f, \hat{\mathbf{n}}) \right] \, , \label{eq:absorption_rate_medium}
\end{align}
where $\Pi_{\hat{h}\hat{h}}(f, \hat{\mathbf{n}})$ is from the graviton-like term in the diagonalized Lagrangian, $\mathcal{L} \supset - \sum_\lambda \hat{h}_{\lambda} \Pi_{\hat{h}\hat{h}}^\lambda \hat{h}_{\lambda}/2$. To compute the absorption we must diagonalize Eq.~\eqref{eq:mixed_Lagrangian} to find $\Pi_{\hat{h}\hat{h}}^\lambda(f, \hat{\mathbf{n}})$.

Gauge freedom allows for immediate simplification in both the photon and graviton sectors, reducing the degrees of freedom from 10 and 4 to 6 and 3 for the graviton and photon field, respectively. We are focused on computing the absorption rate of an incoming vacuum graviton. Since any polarization mixing introduced by the medium is gravitationally suppressed, we can reduce the graviton system to just the two vacuum graviton modes. Projecting the graviton field into the reduced polarization basis as $h_{ij} = e^\lambda_{ij} h_\lambda$, the Lagrangian in Eq.~\eqref{eq:mixed_Lagrangian} simplifies to
\begin{align}
    \mathcal{L} = \frac{1}{2} \begin{pmatrix} h_\lambda & A_\rho \end{pmatrix} 
        \begin{pmatrix} \left( \partial^2 - \Pi_{hh}^\lambda \right) \delta^{\lambda \lambda'} & -\Pi_{h A}^{\lambda,\, \rho'} \\ -\Pi_{A h}^{\rho,\,\lambda'} & \left( \partial^2 \eta^{\rho \rho'} - \partial^\rho \partial^{\rho'} \right) - \Pi_{AA}^{\rho\rho'} \end{pmatrix}
    \begin{pmatrix} h_{\lambda'} \\ A_{\rho'} \end{pmatrix} \label{eq:mixed_step_1}
\end{align}
where $\lambda,\lambda'$ index the usual $+, \times$ graviton polarizations, and self-energies with polarization indices indicate projection onto the polarization vectors, e.g., $\Pi^\lambda_{hh} \equiv e_{ij}^\lambda \Pi_{hh}^{ij, i'j'} e_{i'j'}^\lambda$.

Moving to the photon sector, DM absorption calculations are typically performed in the Lorenz gauge, $\partial_\mu A^\mu = 0$. However, subtleties arise in computing the Feynman diagrams in Eq.~\eqref{eq:example_feynman_diagram} for graviton kinematics, $Q^2 = 0$, since the standard choice of the longitudinal polarization vector, $e_L^\mu = (\omega, \omega \hat{\mathbf{n}}) / \sqrt{-Q^2}$,
is singular. Therefore we work in the Coloumb gauge, $\partial^i A_i = 0$, where the polarization vectors are $e_L^\mu = (1, 0, 0, 0)$, $e_\pm^\mu = (0, \hat{\mathbf{n}}_{\pm})$, and $\hat{\mathbf{n}}_{\pm}$ are two vectors orthonormal to $\hat{\mathbf{n}}$. Projecting the photon field into these polarizations, $A_\mu = e_{\mu}^\lambda A_\lambda$, the Lagrangian in Eq.~\eqref{eq:mixed_step_1} becomes, 
\begin{align}
    \mathcal{L} = -\frac{1}{2}
        \begin{pmatrix} h_\lambda & A_L & A_\sigma \end{pmatrix}
            \begin{pmatrix} 
                \left( -\partial^2 + \Pi_{hh}^\lambda \right) \delta^{\lambda\lambda'} & \Pi_{hL}^\lambda & \Pi_{hA}^{\lambda \sigma'} \\
                \Pi_{L h}^{\lambda'} & -\omega^2 + \Pi_L & \Pi_{LA}^{\sigma'} \\
                \Pi_{Ah}^{\sigma \lambda'} & \Pi_{AL}^\sigma & -\partial^2 \delta^{\sigma \sigma'} + \Pi_{AA}^{\sigma\sigma'}
            \end{pmatrix}
        \begin{pmatrix} h_{\lambda'} \\ A_L \\ A_{\sigma'} \end{pmatrix} \, , \label{eq:mixed_step_2}
\end{align}
where we have explicitly separated the longitudinal photon mode $A_L$ from the transverse modes $A_\sigma$, where $\sigma$ indexes transverse polarizations, $\pm$, and polarization indices represent projections onto polarizations. Projection on to the longitudinal mode is equivalent to projection on to $\hat{n}^i$ by the Ward identity, $Q_\mu \Pi^{\mu \nu}_{AA} = 0$: $\Pi_L \equiv \Pi_{AA}^{00} = \hat{n}^i \Pi_{AA}^{ij} \hat{n}^j$, $\Pi_{AA}^{0\sigma} \equiv \Pi_{LA}^\sigma = \hat{n}^i \Pi_{AA}^{ij} e_{j}^\sigma$, and $\Pi_{AA}^{\sigma 0} \equiv \Pi_{AL}^\sigma = e_i^\sigma \Pi_{AA}^{ij} \hat{n}^j$. 

The Lagrangian in Eq.~\eqref{eq:mixed_step_2} can now be perturbatively diagonalized since the off-diagonal components mixing $h$ and $A$ are $\mathcal{O}(\sqrt{G})$, and therefore $\Pi_{\hat{h}\hat{h}}^\lambda$ is given by,
\begin{align}
    \Pi_{\hat{h}\hat{h}}^\lambda & \approx \Pi_{hh}^\lambda - \Pi_{hA}^{\lambda \alpha} \, \left[ \Delta^{-1}_{A} \right]_{\alpha \alpha'} \, \Pi_{Ah}^{\alpha' \lambda} \label{eq:Pi_hhat_hhat}\\ 
    \Delta_{A}^{\alpha \alpha'} & \equiv 
        \begin{pmatrix}
            -\omega^2 + \Pi_L & \Pi_{LA}^{\sigma'} \\ \Pi_{AL}^\sigma & Q^2 \delta^{\sigma \sigma'} + \Pi_{AA}^{\sigma\sigma'}
        \end{pmatrix}
        \, ,
\end{align}
where $\alpha,\alpha' \in \{ L, \pm \}$ index all the photon polarizations,  and the inverse in Eq.~\eqref{eq:Pi_hhat_hhat} represents the matrix inverse. Note that for an isotropic medium with $\Pi_{AL}^\sigma = \Pi_{LA}^\sigma = 0$, $\Delta_{AA}$ is diagonal, and the inverse in Eq.~\eqref{eq:Pi_hhat_hhat} is trivial. The absorption rate is therefore given by
\begin{align}
    \Gamma_{\lambda} = - \frac{1}{\omega} \text{Im} \left[ \Pi_{hh}^\lambda - \Pi_{hA}^{\lambda \alpha} \, \left[ \Delta^{-1}_{A} \right]_{\alpha \alpha'} \, \Pi_{Ah}^{\alpha' \lambda} \right]  \, . \label{eq:abs_rate_per_graviton}
\end{align}

To compute the self-energies in Eq.~\eqref{eq:abs_rate_per_graviton} we follow Ref.~\cite{Mitridate:2023izi}. As mentioned earlier, these self-energies will generally receive contributions from both the phonons and electrons in the target. However, since our focus is on single-phonon excitations, some simplifications can be made. Notice that the dependence of the absorption rate, $\Gamma_\lambda$, on $\Pi_{hh}$ is only through the imaginary part. Imaginary contributions to self-energies correspond to physical degrees of freedom going on-shell, and therefore below the electronic band gap, $\text{Im} \left[ \Pi_{hh} \right]$ will only receive a contribution from phonon excitations. Furthermore, we expect the electronic contribution to $\Pi_{hA}$ to be sub-dominant, since any GW-electron coupling will be suppressed by $m_e/m_j \sim 10^{-4}$ relative to the GW-ion coupling. Therefore when computing both $\Pi_{hh}$ and $\Pi_{hA}$ we include only the phonon contribution. Calculation of $\Pi_{AA}$ is identical to that described in Ref.~\cite{Mitridate:2023izi}, and will include both electron and phonon contributions.

The single-phonon contribution to the self-energies can be written as~\cite{Mitridate:2023izi}
\begin{align}
    \Pi_{\Phi \Phi'}(f, \hat{\mathbf{n}}) & = - i \sum_\nu \frac{D_\nu(\omega)}{\Omega} \left( \sum_j \bm{\mathcal{F}}_{\Phi, j} \cdot \mathbf{T}_{j \nu \mathbf{q}} \right) \left( \sum_j \bm{\mathcal{F}}_{\Phi', j} \cdot \mathbf{T}_{j \nu\mathbf{q}} \right)^*, \label{eq:phonon_self_energy}\\ 
    D_\nu(\omega) & = \frac{2 i \omega_{\nu}}{\omega^2 - \omega_\nu^2 + i \omega \gamma_\nu} \, , \label{eq:phonon_propagator}
\end{align}
where $\Pi_{\Phi\Phi'}$ implicitly carries the Lorentz indices of the fields $\Phi$ and  $\Phi'$, $D_{\nu}$ is the phonon propagator, and $\mathcal{F}_{\Phi, j}$ is a form factor encapsulating the coupling of the $j^\text{th}$ ion to the field $\Phi$ (and also carries an implicit Lorentz index). The photon form factor is simply the charge of a given ion, and the graviton form factor can be read off from the interaction Hamiltonian in Eq.~\eqref{eq:GW_phonon_interaction_Hamiltonian},
\begin{align}
    \mathcal{F}_{A, j}^{i, \, k} & = e \, \omega \, Q_j \, \delta^{ik}, \label{eq:photon_form_factor} \\ 
    \mathcal{F}_{h, j}^{a i, \, k} & =  - \sqrt{4 \pi G} \, \omega^2 \, m_j  \,  x_{j}^{0, a} \, \delta^{ik} \, , \label{eq:graviton_form_factor}
\end{align}
where $Q_j = N_{p, j} - N_{e, j}$, and $N_{p (e), j}$ are the number of protons (electrons) at site $j$. Substituting Eqs.~\eqref{eq:photon_form_factor} and~\eqref{eq:graviton_form_factor} into Eq.~\eqref{eq:phonon_self_energy} gives
\begin{align}
    \Pi_{hh}^{ik, \, i'k'}(f, \hat{\mathbf{n}}) & = -i \, 4 \pi G \, \omega^4 \sum_\nu \frac{D_\nu(\omega)}{\Omega} \left( \sum_j m_j \, x^{0,\, i}_j \, T^k_{j\nu \mathbf{q}} \right) \left( \sum_j m_j \, x^{0,\, i'}_j \, T^{k'}_{j \nu \mathbf{q}} \right)^*, \\ 
    \Pi_{hA}^{ik, \, i'}(f, \hat{\mathbf{n}}) & = i \, \sqrt{4 \pi G} \, e \, \omega^3 \sum_\nu \frac{D_\nu(\omega)}{\Omega} \left( \sum_j m_j \, x^{0,\, i}_j \, T^k_{j \nu \mathbf{q}} \right) \left( \sum_j Q_j \, T^{i'}_{j \nu \mathbf{q}} \right)^* \, ,
\end{align}
which can then be straightforwardly substituted into Eq.~\eqref{eq:abs_rate_per_graviton} to compute the absorption rate. Note that if the mixing is removed by setting $\Pi_{hA}= \Pi_{Ah} = 0$ (which can be accomplished by setting $Q_j = 0$ as for a non-polar target), and the phonon is assumed to be a perfect resonance, $\gamma \rightarrow 0$, then we recover the absorption rate previously derived from Fermi's Golden Rule in Sec.~\ref{subsubsec:non_polar_targets}. The main effect of the mixing contribution to Eq.\eqref{eq:abs_rate_per_graviton} is to slightly shift the resonance locations from their value in the limit of no mixing, i.e., $\omega_\nu$.

\section{Sensitivity To Gravitational Waves}
\label{sec:results}

With the calculation of the GW absorption rate in Sec.~\ref{sec:theory_derivation}, we can now determine the experimental sensitivity. In the context of DM direct detection, the sensitivity is governed by the number of phonons produced, $N_\text{ph}$, over the observation time, $T$. Requiring that $N_\text{ph} > 3$ is then directly converted to a 95\% C.L. limit on the DM coupling parameters, assuming negligible backgrounds. While $N_\text{ph}$ will still be the figure of merit here, the question of experimental sensitivity becomes more subtle for GW searches since the incoming GWs need not be coherent over $T$, nor have a monochromatic frequency spectrum. These two features are dramatically different than the signal due to absorption of non-relativistic DM, which occurs at frequencies equal to the DM mass and persists over the whole observation time. Due to the relativistic kinematics of the incoming GW, a closer analogy would be to a signal produced by a thermal cosmic axion background~\cite{Dror:2021nyr}, or from DM produced in the Sun~\cite{Derevianko:2010kz,Redondo:2013wwa}. Here, computing the number of phonons produced becomes intimately tied to the parameterization of the signal. Therefore for ease of comparison it is useful to create classes of signals with similar parameterizations. The sources of high-frequency GWs in this frequency range we focus on, for reasons discussed further in Sec.~\ref{subsec:mono_sources}, may be referred to as \textit{deterministic} signals, characterized by a known signal profile which may be parameterized as,\footnote{Deterministic signals may also contain information about direction from which they were emitted. This goes beyond our isotropic assumption in Eq.~\eqref{eq:isotropic_independent_GW_approximation}, and we leave further discussion for future work.}
\begin{align}
    h(t) & = h_0(t) \, e^{2\pi i \,f_s(t) \, t} \, . \label{eq:deterministic_signal}
\end{align}
Stochastic sources, characterized by their power spectral density, are considered in App.~\ref{app:sto_constraints}. In Eq.~\eqref{eq:deterministic_signal}, $f_s(t)$ is the instantaneous signal frequency at time $t$, and $h_0(t)$ is the signal amplitude at time $t$. Assuming that $f_s$ and $h_0$ are slowly-varying functions of time, the GW energy density for deterministic signals is
\begin{align}
    \rho_\text{GW}(t) = \frac{\pi}{8G } \, f_s^2(t) \, h_0^2(t)  \equiv \rho_\text{GW}^0(f_s) \, h_0^2(t)\, , \label{eq:rho_GW}
\end{align}
where we have defined, $\rho_\text{GW}^0(f) \equiv (\pi/8) f^2 / G$. The GW energy density parameter $\Omega_\text{GW}$ for deterministic signals is then given by
\begin{align}
    \Omega_\text{GW}(f, t) = \frac{\rho_\text{GW}^0(f)}{ \rho_c} \,  f \, h_0^2(t) \, \delta( f - f_s ) \, .
    \label{eq:Omega_GW_mono}
\end{align}
Note that while $\Omega_\text{GW}(f)$ is typically used in the context of stochastic GW signals, its use here can be understood as just a rewriting of the differential energy density, $d \rho_\text{GW} / d f$, given by Eq.~\eqref{eq:isotropic_independent_GW_approximation}; $\rho_\text{GW}$ in Eq.~\eqref{eq:rho_GW} is related to $\Omega_\text{GW}$ in Eq.~\eqref{eq:Omega_GW_mono} by $\rho_\text{GW} = \int (d \rho_\text{GW} / df) \, df = \rho_c \int \Omega_\text{GW}(f) \, d \ln{f}$. The expected number of phonons produced during $T$ is found by integrating the rate in Eq.~\eqref{eq:total_averaged_GW_abs_rate}, and multiplying by the detector mass, $M$,
\begin{align}
    N_\text{ph} = M \int_0^T dt \int df \frac{1}{4 \pi f} \frac{\rho_\text{GW}^0(f)}{\rho_T} \, h_0^2(t) \, \Gamma(f) \, \delta(f - f_s(t)) \, . \label{eq:n_phonon_produced}
\end{align}

While Eq.~\eqref{eq:n_phonon_produced} is the underlying quantity which determines the sensitivity, comparing the response to different deterministic signals, for example those with different time dependence, is subtle. Ideally, the signal strength and detector sensitivity would be completely decoupled, such that if the signal strength is greater than the detector sensitivity, then the signal may be seen. This decomposition is most clear in the context of \textit{monochromatic} signals, for which $f_s$ is constant, and the amplitude is slowly varying. For these signals, we can define,
\begin{align}
    h_s^2 & \equiv \frac{1}{T} \int_0^T  h_0^2(t) \, dt  \label{eq:h_s_mono}\, ,\\ 
    h_\text{det}^2(f) & \equiv 3 \; \frac{4 \pi f}{ \rho_\text{GW}^0 \, V \, T \, \Gamma(f)} \label{eq:h_det_definition} \, ,
\end{align}
where $h_s$ characterizes the signal strength and $h_\text{det}$ characterizes the detector sensitivity. This allows Eq.~\eqref{eq:n_phonon_produced} to be written as
\begin{align}
    N_\text{ph} & = 3 \frac{h_s^2}{h_\text{det}^2(f_s)} \, . \label{eq:n_phonon_monochromatic_deterministic}
\end{align}
Therefore when $h_s^2 > h_\text{det}^2(f_s)$ the number of phonons produced is greater than $3$, and limits on the signal can be set at the 95\% C.L.

There is redundancy in our definitions of $h_s$ and $h_\text{det}(f)$, since only their relative magnitude is important, and therefore great care should be taken to compare the sensitivity of different experiments which may differ in definitions of $h_s$ and $h_\text{det}$. The definitions chosen here are useful because they do not only apply to detection via single-phonon excitations. For any direct detection experiment where the primary observable is some number of excitations, and the interaction of a GW with that excitation is $\Gamma(f)$, Eq.~\eqref{eq:h_det_definition} can be used to define the detector sensitivity.

Another type of deterministic signal is a \textit{chirp} signal, where $\dot{f}_s = df_s/dt$ is known. For these signals the time dependence in $h_0(t)$ can be traded for frequency dependence by solving for $f = f_s(t)$. Defining the signal strength in this case as
\begin{align}
    h_s^2(f) & \equiv \frac{f}{\dot{f} \, T} h_0^2(f) = \frac{\tau_f}{T} \, h_0^2(f) \, , \label{eq:h_s_chirp}
\end{align}
where $\tau_f \equiv f_s / \dot{f}_s$ is the signal coherence time, the number of phonons produced from this signal type is can then be calculated, using $dt = df / \dot{f}$, as
\begin{align}
    N_\text{ph} & = 3 \int \, \frac{h_s^2(f)}{h_\text{det}^2(f)} \, d\ln{f}  \, , \label{eq:N_ph_chirp}
\end{align}
where $h_\text{det}$ has the same definition as Eq.~\eqref{eq:h_det_definition}, and the frequency integral is performed over $f_s(0) \leq f \leq f_s(T)$. We see that if $h_s \gtrsim h_\text{det}$ over an e-fold in $f$, then the number of phonons generated will be larger than $3$. 

With Eq.~\eqref{eq:h_det_definition}, the detector sensitivity, $h_\text{det}$, can now be computed in a general target material. In this section we focus on nine targets that have been previously studied in the context of DM direct detection. \ce{GaAs} and \ce{Al2O3} are being utilized in the TESSARACT experiment~\cite{Chang2020}, specifically designed to search for DM-induced single-phonon excitations, and \ce{Si} is used in SuperCDMS CPD~\cite{SuperCDMS:2020aus} which is sensitive to secondary athermal phonons produced from a DM-induced event. \ce{Si} and \ce{Ge} have been used in other experiments, e.g., CDEX~\cite{CDEX:2022kcd}, DAMIC~\cite{DAMIC:2015znm,DAMIC:2016qck,DAMIC:2019dcn,DAMIC:2020cut,Settimo:2020cbq}, EDELWEISS~\cite{EDELWEISS:2018tde,EDELWEISS:2019vjv,EDELWEISS:2020fxc}, SENSEI~\cite{Crisler:2018gci,SENSEI:2019ibb,SENSEI:2020dpa}, and SuperCDMS~\cite{CDMS:2009fba,SuperCDMS:2018mne,SuperCDMS:2020ymb}, searching for DM-induced electronic excitations and nuclear recoils. \ce{NaI} (DAMA/LIBRA~\cite{Baum:2018ekm}, KIMS~\cite{Kim:2015prm}, ANAIS~\cite{Amare:2019jul}, SABRE~\cite{Shields:2015wka}, DM-Ice~\cite{Jo:2016qql}), \ce{CsI} (KIMS~\cite{Kim:2008zzn}), and \ce{CaWO4} (CRESST~\cite{Probst:2002qb,CRESST:2015txj, CRESST:2017cdd}) have also been used as target material in nuclear recoil experiments. In addition to these we include two more speculative targets: \ce{SiO2}, shown to have strong phonon responses to absorption and scattering of dark photon DM~\cite{Griffin:2019mvc,Trickle:2020oki,Mitridate:2020kly, Coskuner:2021qxo,Mitridate:2023izi}, and diamond, which has been independently proposed to search for electronic excitations and nuclear recoils~\cite{Kurinsky:2019pgb,Canonica:2020omq}.

While it is likely not all of these targets will be employed as single-phonon detectors, comparing the projected sensitivity for a variety of targets illustrates how they can complement each other. The number of gapped phonon modes, and their energy spectrum, varies across targets. For example, \ce{CsI} has a single resonance at $\omega \sim 10 \, \text{meV}$, diamond has one at $\omega \sim 175 \, \text{meV}$, and \ce{CaWO4} has many resonances between $8 \, \text{meV}$ and $110 \, \text{meV}$. The total number of gapped modes is $3n - 3$, where $n$ is the number of atoms in the unit cell, and the energy of the lowest gapped mode correlates with the speed of sound, which varies across materials. The GW sensitivity will be peaked near these resonance frequencies, and therefore a judicious choice of targets is necessary to cover the broadest possible frequency range. See Ref.~\cite{Griffin:2019mvc} for the phonon band structures for the targets considered here. Additionally, in contrast to the search for dark photon DM~\cite{Knapen:2017ekk,Griffin:2018bjn,Knapen:2021bwg,Mitridate:2023izi}, or axion DM in a background magnetic field~\cite{Mitridate:2020kly,Berlin:2023ppd} the targets do \textit{not} have to be polar, i.e., contain oppositely charged ions in the unit cell. This allows for targets like diamond, Si, and Ge to be used to search for GWs even though they are not useful in searching for specific benchmark sub-eV DM models, which instead must rely on multiphonon processes~\cite{Knapen:2021bwg} at $\mathcal{O}(\text{meV})$ frequencies.

\begin{figure}[t]
    \centering
    \includegraphics[width=\textwidth]{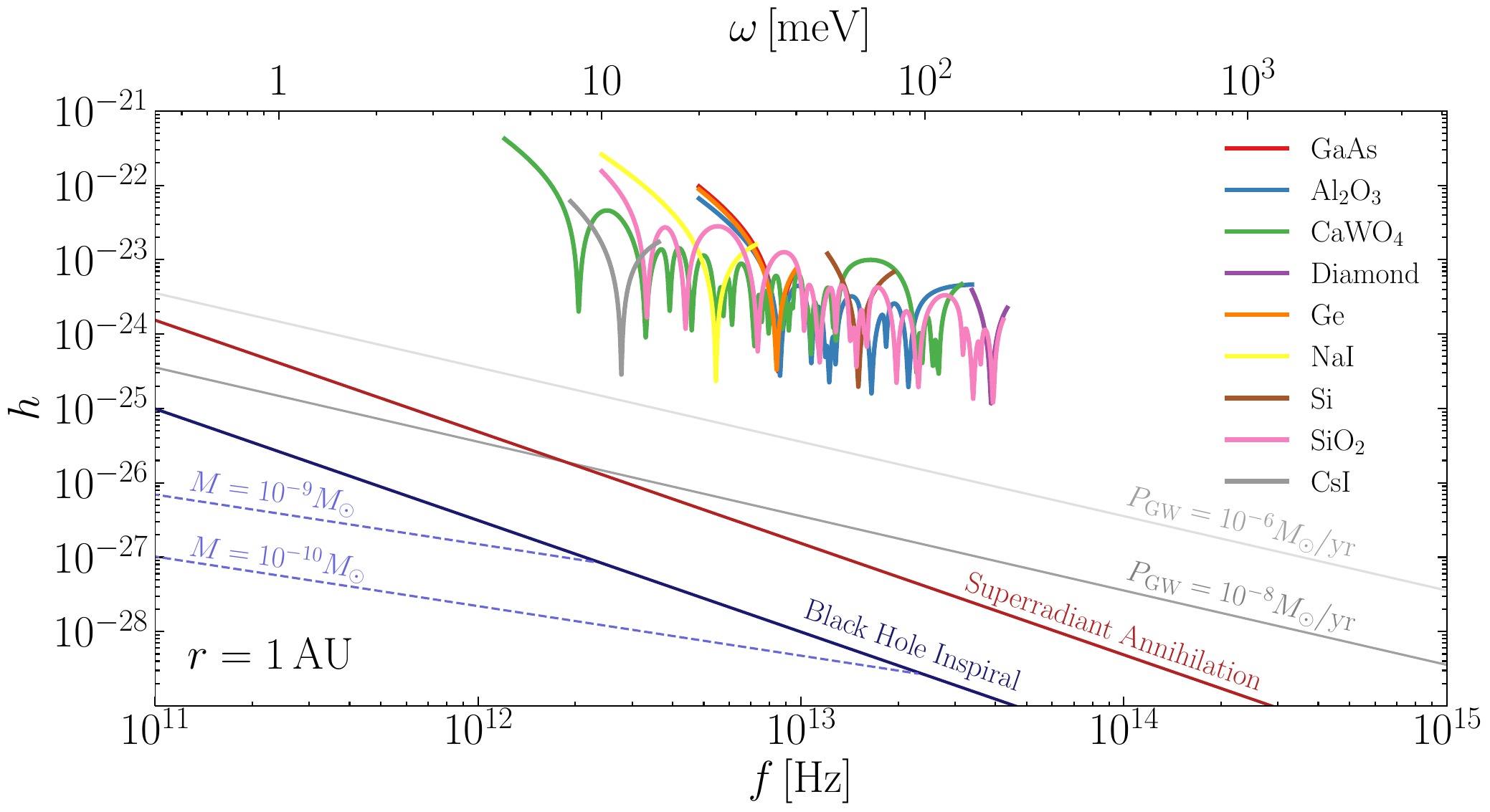}
    \caption{Detector sensitivities, $h_\text{det}$, Eq.~\eqref{eq:h_det_definition}, for experiments utilizing gapped, single-phonon excitations in a variety of crystal targets, assuming negligible backgrounds and a $\text{kg} \cdot \text{yr}$ exposure. The frequency, and number, of resonances are properties of the specific target. All line widths, $\gamma_\nu$, Eq.~\eqref{eq:phonon_propagator}, are taken to be $10^{-2} \omega_\nu$, where $\omega_\nu$ is the phonon frequency. The projected sensitivity is cut off away from the peak resonances for visual simplicity, and to emphasize that the simplistic modelling of $\gamma_\nu$ discussed previously is unlikely to hold far from the resonances. Note that Si, Ge, and Diamond are non-polar, whereas the rest of the targets are polar. In addition to the detector sensitivity we show the signal strength, $h_s$, for three sources of high-frequency GWs, all assuming sources $r = 1\,\text{AU}$ away and $T = 1\,\text{yr}$. Gray lines represent idealized, benchmark sources, computed with Eq.~\eqref{eq:h0_any_source}, which emit GWs with power $P_\text{GW}$, and are coherent over the observation time, i.e., monochromatic sources with $h_s \approx h_0$ in Eq.~\eqref{eq:h_s_mono}. The red line, labelled ``Superradiant Annihilation", is a monochromatic signal, from the annihilation of two bosons in a superradiant produced population around low-mass ($M \lesssim 10^{-7}\,M_\odot$) BHs. The signal strength is computed with Eq.~\eqref{eq:h_s_sr_ann}, assuming the BH mass is maximal at each frequency; see Sec.~\ref{subsubsec:superradiance} for more details. The blue lines, labelled ``Black Hole Inspiral", are chirp signals from the inspiral of two low-mass ($M \lesssim 10^{-8}\,M_\odot$) BHs. The light blue dashed lines correspond to $h_s$ in Eq.~\eqref{eq:h_s_bhm} for fixed BH mass, $M$. The solid line represents the high-frequency boundary of possible BH inspiral signal strengths; at each $f$, we choose the BH mass $M$ in Eq.~\eqref{eq:h_s_bhm} whose $f_\text{ISCO}(M) = f$.}
    \label{fig:monochromatic_detail_projections}
\end{figure}

Understanding and mitigating backgrounds is crucial to the success of any direct detection experiment utilizing single-phonon excitations. The dominant irreducible background is from coherent solar neutrino scattering; the cosmic neutrino background is negligible. Solar neutrinos are expected to produce a background of phonons at a rate of $R \sim 10^{-2} / \text{meV} / \text{kg} \cdot \text{yr}$ between $1\, \text{meV} \lesssim \omega \lesssim 100 \, \text{meV}$~\cite{Hochberg:2015fth,Berghaus:2021wrp}. Other backgrounds are expected to be important, and in fact dominate current low threshold direct detection technology. Examples of such backgrounds are those induced by cosmic high energy particles~\cite{Du:2020ldo}, and from radiogenic sources in the detector or shielding~\cite{Berghaus:2021wrp}, the latter generating a background of phonons at a rate of $R \sim 1-10 / \text{meV} / \text{kg} \cdot \text{yr}$ with current levels of radio-purity. Improved shielding, active signal vetoing, and more radio-pure samples will be essential to reduce these backgrounds. Vibrational noise is unlikely to be important since any source frequency is far from the phonon resonances of interest here, and thermal noise will be Boltzmann suppressed. 

In Fig.~\ref{fig:monochromatic_detail_projections} we compare the detector sensitivity, $h_\text{det}$, of the nine target materials assuming a $\text{kg} \cdot \text{yr}$ exposure and backgrounds at the irreducible level, $\lesssim 1/\text{kg} \cdot \text{yr}$, which are negligible. The phonon line widths are taken to be $\gamma_\nu = 10^{-2} \, \omega_\nu$, values which have been shown to reproduce the dielectric function reasonably well~\cite{Mitridate:2023izi}. In addition to comparing the sensitivity of different targets, we also compare $h_\text{det}$ to the signal strength, $h_s$, for the sources discussed in Sec.~\ref{subsec:mono_sources}, all assumed to be $r = 1\,\text{AU}$ away. Gray lines are shown as benchmark lines and correspond to idealized, monochromatic sources which emit GW radiation from a single source isotropically with power $P_\text{GW}$ over the observation time. The sensitivity could be further improved by increasing the detector volume, $h_\text{det} \propto 1/\sqrt{V}$, as well as finding targets with smaller phonon linewidths, $\gamma_\nu$, since at the phonon peak, $h_\text{det}$ is proportional to $\sqrt{\gamma_\nu}$. 

\begin{figure}[t!]
    \centering
    \includegraphics[width=\textwidth]{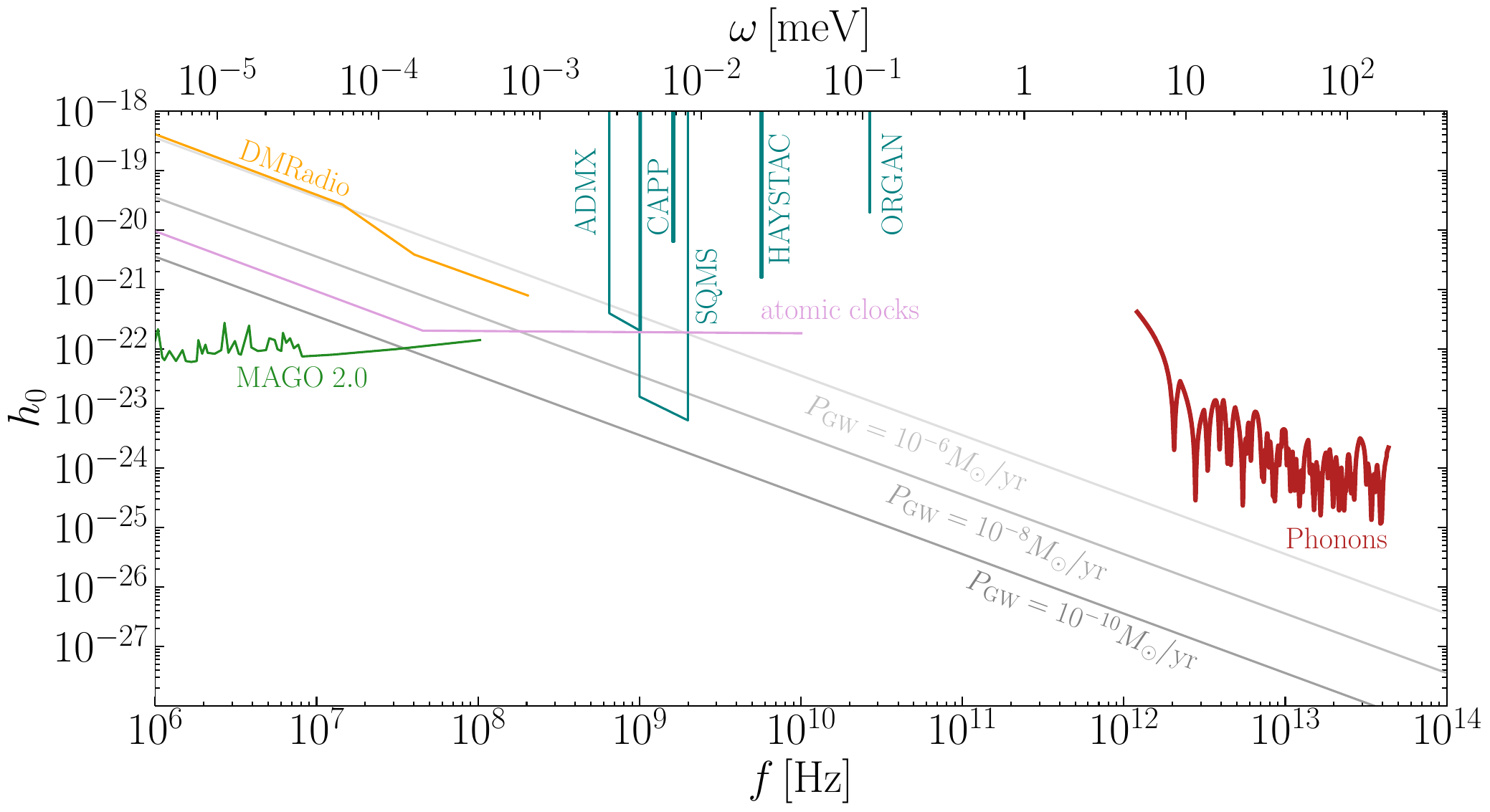}
    \caption{Comparing the sensitivity of high-frequency GW detectors in the $\text{MHz} \lesssim f \lesssim 100 \, \text{THz}$ range. The comparison between detectors is only meaningful for monochromatic signals with constant amplitude, i.e., those of the form $h(t) \sim h_0 \, e^{2 \pi i f t}$, where $h_0$ is a constant. For these signals, if the $h_0$ produced from a source is greater than the detector sensitivity, then it may be detected. The gray lines correspond to example, benchmark, sources isotropically emitting GWs with power $P_\text{GW}$, at frequency $f$, and at a distance of $r = 1\,\text{AU}$. $h_0$ is computed using Eq.~\eqref{eq:h0_any_source}. The solid red line labelled ``Phonons" is an outline of the specific target sensitivities shown in Fig.~\ref{fig:monochromatic_detail_projections}, which assume $T = 1\,\text{yr}$. The sensitivity of different microwave cavities from Ref.~\cite{Berlin:2021txa} (ADMX~\cite{ADMX:2018ogs, ADMX:2019uok, ADMX:2021nhd}, CAPP~\cite{Lee:2020cfj}, HAYSTAC~\cite{HAYSTAC:2018rwy}, ORGAN~\cite{McAllister:2017lkb}, and SQMS~\cite{Berlin:2021txa}) is shown in teal. The sensitivity of the MAGO 2.0 proposal from Ref.~\cite{Berlin:2023grv} are shown in green. The sensitivities of DMRadio-GUT~\cite{DMRadio:2022jfv} and DMRadio-$\text{m}^3$~\cite{DMRadio:2022pkf}, assuming a figure-8 pickup loop, have been combined, labelled ``DMRadio", from Ref.~\cite{Domcke:2023bat} and are shown in orange. The sensitivity of atomic clocks are from Ref.~\cite{Bringmann:2023gba} and shown in purple.}
    \label{fig:monochromatic_overview_projections}
\end{figure}

In Fig.~\ref{fig:monochromatic_overview_projections} we compare the sensitivity of single-phonon detectors to other high-frequency GW detectors in the $\text{MHz} \lesssim f \lesssim 100 \, \text{THz}$ frequency range. The curve labelled ``Phonons" is simply an outline of the sensitivities in Fig.~\ref{fig:monochromatic_detail_projections}. The detectors are compared in their ability to detect monochromatic GWs, coherent over all their observation times, with (time-independent) amplitude $h_0$. Constraints from recasting limits from current experimental data (ADMX~\cite{ADMX:2018ogs, ADMX:2019uok, ADMX:2021nhd}, CAPP~\cite{Lee:2020cfj}, HAYSTAC~\cite{HAYSTAC:2018rwy}, ORGAN~\cite{McAllister:2017lkb}) were taken directly from Ref.~\cite{Berlin:2021txa}. Projections for other experiments have been rescaled assuming a run time, $T = 1 \, \text{yr}$. This changes the projections for a future detector built in the SQMS Center at Fermilab~\cite{Berlin:2021txa}, MAGO 2.0~\cite{Berlin:2023grv}, DMRadio (a combination of projections from DMRadio-GUT~\cite{DMRadio:2022jfv} and DMRadio-m$^3$~\cite{DMRadio:2022pkf} from Ref.~\cite{Domcke:2023bat} assuming figure-8 pickup loop), and the ``realistic" setup for the atomic clock experiment from Ref.~\cite{Bringmann:2023gba} by $\mathcal{O}(1)$ factors relative to their respective references. The sensitivity of other axion haloscopes, ABRACADABRA~\cite{Kahn:2016aff,Ouellet:2018beu,Salemi:2021gck}, ADMX SLIC~\cite{Crisosto:2019fcj}, BASE~\cite{Devlin:2021fpq}, SHAFT~\cite{Gramolin:2020ict}, WISPLC~\cite{Zhang:2021bpa}, can be found in Ref.~\cite{Domcke:2023bat}, and are too weak to appear in Fig.~\ref{fig:monochromatic_overview_projections}. 

\section{Sources}
\label{subsec:mono_sources}

\begin{figure}[t]
    \centering
    \includegraphics[width=\textwidth]{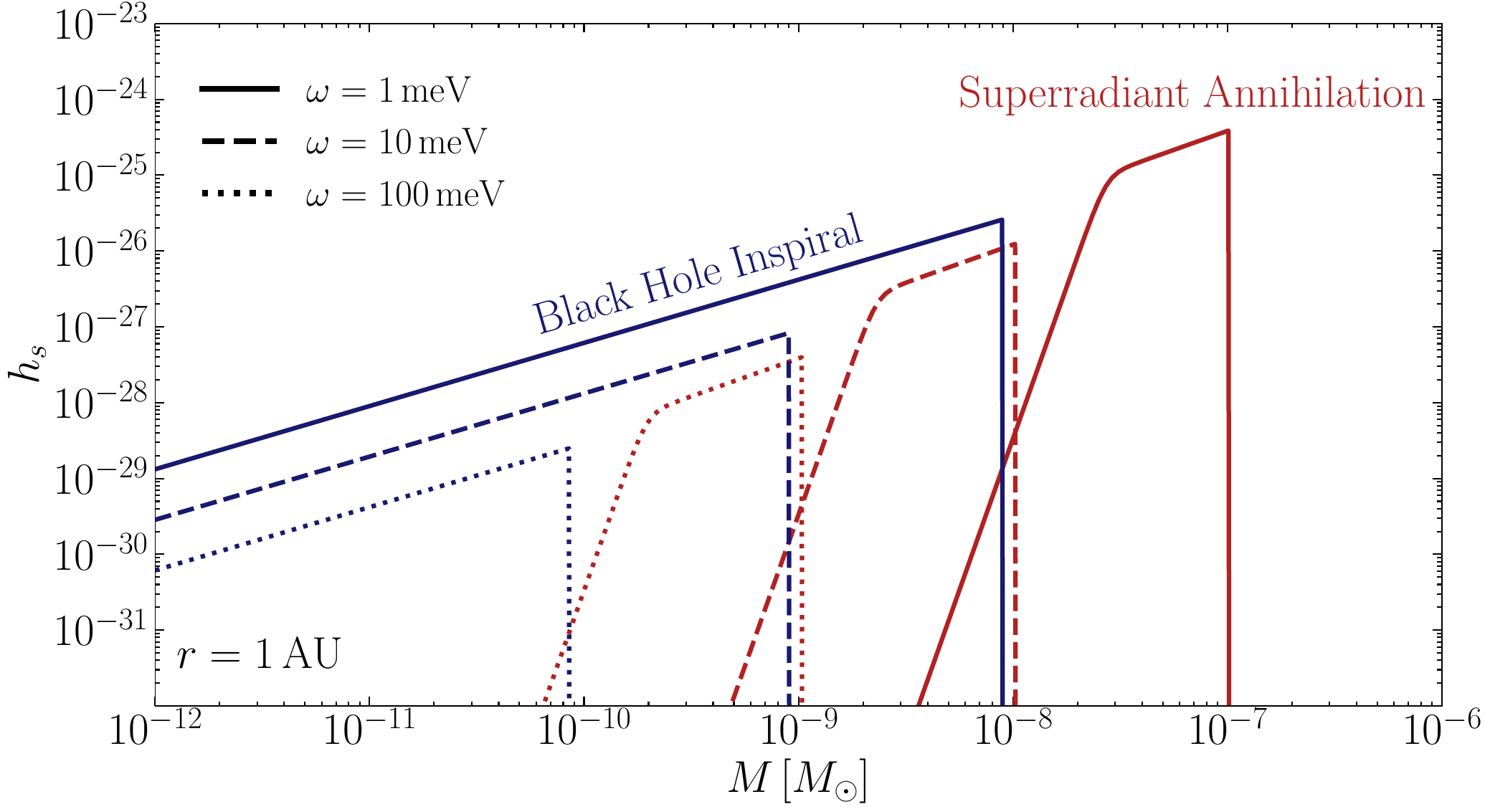}
    \caption{Comparison of the signal strength, $h_s$, from the inspiral of BHs with equal mass $M$ (``Black Hole Inspiral", blue, Sec.~\ref{subsubsec:BH_merger}), and superradiant annihilation of bosons surrounding a BH of mass $M$ (``Superradiant Annihilation", red, Sec.~\ref{subsubsec:superradiance}) at different GW frequencies, $\omega$. All lines assume the source is $r = 1 \, \text{AU}$ away and $T = 1\, \text{yr}$. The superradiant annihilation signal is computed with Eq.~\eqref{eq:h_s_sr_ann}. The turnover occurs when the superradiant annihilation timescale, Eq.~\eqref{eq:tau_sr_ann}, is equal to the observation time, $T$, and the cutoff is from requiring $G M \omega < 0.8$, corresponding to the point where numerical simulations indicate a breakdown of our analytic approximations. The BH inspiral signal is shown at the different points in its frequency evolution and computed with Eq.~\eqref{eq:h_s_bhm}. The cutoff occurs for BH masses too large to emit GWs at $\omega$, i.e., their $\omega_{\text{ISCO}} < \omega$.
    \label{fig:h_s_comparison}}
\end{figure}

The universe naturally provides a variety of GW sources. The Standard Model predicts GW production over a range of frequencies, from the inspiral of supermassive black hole binaries~\cite{Rajagopal:1994zj,Jaffe:2002rt,Wyithe_2003,Sesana:2004sp,Burke-Spolaor:2018bvk} at $f \sim \text{nHz}$, to populations of gravitons frozen in from the early universe at $f \sim \text{THz}$~\cite{Ghiglieri:2015nfa,Ghiglieri:2020mhm,Ringwald:2020ist,Ghiglieri:2022rfp,Muia:2023wru}. Additionally, many theories beyond the Standard Model also predict GW production from inflation~\cite{Grishchuk:1974ny,Starobinsky:1979ty,Rubakov:1982df,Fabbri:1983us,Abbott:1984fp,Domenech:2021ztg}, inflaton annihilation into gravitons~\cite{Ema:2015dka,Ema:2016hlw,Ema:2020ggo}, phase transitions~\cite{Witten:1984rs,Hogan:1986qda,Kosowsky:1991ua,Kosowsky:1992vn,Kosowsky:1992rz,Kamionkowski:1993fg,Caprini:2007xq,Huber:2008hg,Hindmarsh:2013xza,Giblin_2014,Hindmarsh:2015qta}, preheating~\cite{Khlebnikov:1997di,Lozanov}, topological defects~\cite{Kibble:1976sj,Damour:2000wa,Damour:2001bk}, noisy turbulent motion~\cite{Kosowsky:2001xp,Nicolis:2003tg,Caprini:2006jb,Gogoberidze:2007an,Kalaydzhyan:2014wca} and equilibriated gravitons~\cite{Kolb:206230,Vagnozzi:2022qmc}; see Refs.~\cite{Maggiore:1999vm,Caprini:2018mtu,Christensen_2018} for reviews. 

However, if these GWs are present before big-bang nucleosynthesis (BBN), they would contribute significantly to the radiation density of the universe, which is strongly constrained at that time~\cite{Cyburt:2015mya}. The total amount of GW radiation present at BBN is restricted to be less than $\sim 2 \times 10^{-6}$~\cite{Cyburt:2015mya,Aggarwal:2020olq} of the total energy density in the universe. This severely limits the amplitude of these GWs today, especially at the high frequencies we are concerned with here. The characteristic strain of these cosmological, stochastic GWs is given by~\cite{Moore:2014lga,Aggarwal:2020olq},
\begin{align}
    h_c(f) & \equiv \sqrt{\frac{3 H_0^2}{4 \pi^2} \frac{\Omega_\text{GW}(f)}{f^2}}  \sim 10^{-33} \left( \frac{\text{THz}}{f} \right) \left( \frac{\Omega_\text{GW}(f)}{10^{-6}} \right)^\frac{1}{2} \, ,   
    \label{eq:hc_strain_definition}
\end{align}
which, while not an exact ``apples to apples" comparison (see App.~\ref{app:sto_constraints} for more details), is many orders of magnitude smaller than the detector sensitivities shown in Fig.~\ref{fig:monochromatic_overview_projections}. To emphasize how un-detectable this is, note that the detector sensitivity in Eq.~\eqref{eq:h_det_definition} scales as $L^{-3/2}$, where $L$ is the length scale of the detector. A detector would need to be scaled to $L \sim 100 \, \text{km}$, while still maintaining zero background, to reach $h_c \sim 10^{-33}$ from its sensitivity of $10^{-24}$ at $L \sim 10 \, \text{cm}$. Therefore it will be challenging for detectors based on single-phonon sensitivity to detect these stochastic, early-universe GWs. 

GWs may also be produced on astrophysical scales, i.e., by processes occurring in the Milky Way. Independent of how the GWs are generated, the GW amplitude $h_0$ from any source emitting GWs isotropically with power $P_\text{GW}$ at frequency $\omega$ and a distance $r$ away is given by,
\begin{align}
    h_0 = \sqrt{\frac{8 G P_\text{GW}}{\omega^2 r^2}} \sim 10^{-24} \, \left( \frac{P_\text{GW}}{10^{12} \, M_\odot / \text{yr}} \right)^\frac{1}{2} \left( \frac{1 \, \text{meV}}{\omega} \right) \left( \frac{8 \, \text{kpc}}{r} \right) \, . \label{eq:h0_any_source}
\end{align}
Achieving a detectable strain at galactic distances requires an extraordinary amount of power. Said another way, if in one year the entire mass-energy of the Milky Way ($\sim 10^{12} \, M_\odot$) were converted to $\text{meV}$-frequency GWs at the galactic center ($r = 8 \, \text{kpc}$), the amplitude would be only slightly above the detector sensitivity. 

Therefore it appears the only reasonable source of GWs must come from a source at sub-kpc scales. Normalizing Eq.~\eqref{eq:h0_any_source} to AU distance scales, the required $P_\text{GW}$ is much smaller,
\begin{align}
    h_0 \sim 10^{-24} \, \left( \frac{P_\text{GW}}{10^{-6} \, M_\odot / \text{yr}} \right)^\frac{1}{2} \left( \frac{1 \, \text{meV}}{\omega} \right) \left( \frac{1 \, \text{AU}}{r} \right) \, . \label{eq:h0_any_source_AU}
\end{align}
This signal strength is shown as a gray line in Fig.~\ref{fig:monochromatic_overview_projections}, and is more emblematic of the benchmark signals that very high-frequency GW experiments may be sensitive to.\footnote{Note that if compact objects of mass $M$ constitute all of the local DM then the typical separation is $(\rho_\text{DM} / M)^{-1/3}~\sim~100 \, \text{AU} \left( M / 10^{-12} M_\odot \right)^{1/3}$, for $\rho_\text{DM} = 0.4 \, \text{GeV} / \text{cm}^3$. While the distance scale is much closer to AU scale of interest, the required masses are too small to generate meaningful GW signals.}

We now consider two specific realizations of these deterministic signals: superradiant annihilation of bosons surrounding a BH (Sec.~\ref{subsubsec:superradiance}), and BH inspiral (Sec.~\ref{subsubsec:BH_merger}). Their signal strengths, $h_s$, are shown in Fig.~\ref{fig:h_s_comparison} as a function of the BH mass.

\subsection{Superradiant Annihilation}
\label{subsubsec:superradiance}

Superradiance is the process where a rotating BH of mass $M$ creates a high occupancy bosonic cloud of particles by losing mass and angular momentum~\cite{Arvanitaki:2010sy,Arvanitaki:2014wva,Brito:2015oca,Baryakhtar:2020gao}. The bosons that are produced occupy bound states around the BH, which are analogous to bound states in a hydrogen atom since the gravitational potential is approximately $1/r$ far enough from the BH. The production rate of bosons in a given bound state depends on the number of bosons present, and is therefore exponentially enhanced until the BH is spinning too slowly for superradiance to occur. The maximum number of superradiantly produced bosons in the ``$n = 2,\,\ell = m = 1$" mode, to borrow the notation of atomic physics, is,
\begin{align}
    N_\text{sr} & = \Delta a_{*} \left( \frac{M}{M_\text{pl}} \right)^2 \sim 10^{76} \times \Delta a_{*} \left( \frac{M}{M_\odot} \right)^2 \, .\label{eq:N_sr}
\end{align}
where $M_\text{pl} = 1.22 \times 10^{19} \, \text{GeV}$, and $\Delta a_*$ is the difference in the BH spin parameter, $a_* = J \, (GM^2)^{-1}$, where $J$ is the BH angular momentum, before and after superradiance occurs. These superradiantly produced bosons may then annihilate in to monochromatic GWs at frequencies $\omega = 2 \mu$, where $\mu$ is the mass of the boson.\footnote{Level transitions also produce GWs, although at a subdominant rate to annihilation, and therefore we focus on the annihilation process here.} For superradiance to occur, the BH must be spinning fast enough, $m \Omega_H > \mu$, where $\Omega_H$ is the angular velocity of the BH at the horizon. This is known as the ``superradiance condition" and, when written in terms of the BH mass, requires $G M \omega < 1$ for the $\ell = 1$ mode. Therefore there is a maximum BH mass which can emit GWs at frequencies $\omega$ by superradiant annihilation, $M_\text{max}^{\text{sr}} = M_\text{pl}^2 / \omega \sim 10^{-7} M_\odot \, \left( \text{meV} / \omega \right)$. 

Computing the annihilation rate of two bosons in any state is a difficult problem because the gravitational ``fine structure constant", $\alpha \equiv G M \mu$, can be non-perturbatively large. Therefore while analytic solutions exist for $\alpha \ll 1$, one must resort to numerical relativity simulations~\cite{Dolan:2007mj,Yoshino:2013ofa,East:2017ovw,Witek:2012tr} for $\alpha \lesssim 1/2$, near the edge of the superradiance condition. These simulations indicate that the maximum annihilation signal occurs for $0.25 \lesssim \alpha \lesssim 0.5$~\cite{Yoshino:2013ofa,East:2017ovw}, depending on the spin of the boson and the initial $a_*$ of the BH. For simplicity we adopt a semi-analytic approach to model the annihilation rate similar to Ref.~\cite{Arvanitaki:2014wva}, using the scaling relations from the perturbative regime, with an overall coefficient set by the numerical simulations, and restrict $\alpha < 0.4$. With this parameterization the annihilation rate of two bosons in the $n = 2,\,\ell = m = 1$ state is given by
\begin{align}
    \Gamma_{\text{sr},\,\text{ann}} & = C_\Gamma \, M_\text{pl} \left( 2 \alpha \right)^{15} \left( \frac{M_\text{Pl}}{M} \right)^3 \sim \frac{10^{-54}}{\text{yr}} \left( \frac{M}{10^{-7} \, M_\odot} \right)^{12} \left( \frac{\omega}{\text{meV}} \right)^{15} \, ,\label{eq:sr_ann_rate_per_interaction}
\end{align}
and where the overall coefficient, $C_\Gamma \approx 10^{-10}$, is found from the numerical simulations in Ref.~\cite{Yoshino:2013ofa}. The total annihilation rate is further enhanced by the number of bosons in the cloud,
\begin{align}
    R_{\text{sr},\,\text{ann}} & = N_\text{sr}^2 \, \Gamma_{\text{sr},\,\text{ann}} \sim \frac{10^{70}}{\text{yr}} \, \Delta a_*^2 \left( \frac{M}{10^{-7}\,M_\odot} \right)^{16} \left( \frac{\omega}{\text{meV}} \right)^{15} \, , \label{eq:sr_ann_total_rate}
\end{align}
where the $N_{\rm sr}^2$ factor is due to the fact that the GW is emitted via an annihilation process.
This generates $P_\text{GW} = \omega R_{\text{sr},\,\text{ann}}$ of power in outgoing GWs. The annihilation process continues until all the bosons have annihilated to GWs. The timescale for this to happen is also set by Eq.~\eqref{eq:sr_ann_total_rate},
\begin{align}
    \tau_{\text{sr},\,\text{ann}} & =  \frac{N_\text{sr}}{R_{\text{sr},\,\text{ann}}} \sim  10^{-8} \, \text{yr} \, \frac{1}{\Delta a_*} \left( \frac{10^{-7} \, M_\odot}{M} \right)^{14} \left( \frac{\text{meV}}{\omega} \right)^{15} \, . \label{eq:tau_sr_ann}
\end{align}
The amplitude of the GWs emitted by superradiant annihilation is given by~\cite{Arvanitaki:2014wva}
\begin{align}
    h_0^{\text{sr}, \text{ann}}(t) & = \sqrt{\frac{8 G}{r^2 \omega} R_{\text{sr},\, \text{ann}}} \, \frac{1}{1 + t / \tau_{\text{sr},\,\text{ann}} } \, , \label{eq:h0_sr_ann}
\end{align}
which can then be substituted in to the definition of the signal strength, $h_s$, for monochromatic signals given in Eq.~\eqref{eq:h_s_mono}. The signal strength has parametrically different behavior depending on whether the signal lasts over the entire observation time:
\begin{align}
    h_s^{\text{sr},\,\text{ann}} = 
        \begin{cases} 
            \displaystyle \sqrt{\frac{8 G}{r^2 \omega} R_{\text{sr},\,\text{ann}} } \sim 10^{-21} \, \Delta a_* \left( \frac{\text{AU}}{r} \right) \left( \frac{M}{10^{-7}\, M_\odot} \right)^8 \left( \frac{\omega}{\text{meV}} \right)^7 & \tau_{\text{sr},\,\text{ann}} \gg T \vspace{1em} \\ 
            \displaystyle\sqrt{\frac{8 G}{r^2 \omega} R_{\text{sr},\,\text{ann}} } \sqrt{\frac{\tau_{\text{sr},\,\text{ann}}}{T}} \sim 10^{-25} \, \sqrt{\Delta a_*} \left( \frac{\text{AU}}{r} \right) \left( \frac{\text{yr}}{T} \right)^\frac{1}{2} \left( \frac{M}{10^{-7} \, M_\odot} \right) \left( \frac{\text{meV}}{\omega} \right)^\frac{1}{2} & \tau_{\text{sr},\,\text{ann}} \ll T \, .
        \end{cases} \label{eq:h_s_sr_ann}
\end{align}

Eq.~\eqref{eq:h_s_sr_ann} is shown in Fig.~\ref{fig:h_s_comparison} (``Superradiant Annihilation") as a function of BH mass, for different choices of signal frequency, assuming $T = 1\,\text{yr}$. The maximum BH mass is set by $M_\text{max}^\text{sr}$. The turnover for each curve occurs when $\tau_{\text{sr},\,\text{ann}} \sim T$. BH masses above the turnover have $\tau_{\text{sr},\,\text{ann}} \ll T$, and those below have $\tau_{\text{sr},\,\text{ann}} \gg T$. Lastly, we note that while it may seem beneficial to increase the annihilation rate, $\Gamma_{\text{sr},\,\text{ann}}$ in Eq.~\eqref{eq:sr_ann_rate_per_interaction}, to increase $h_s$ and GW power output, the timescale over which these GWs are emitted decreases, which hinders the signal. In fact, these competing effects exactly compensate each other. If the $\Gamma_{\text{sr},\,\text{ann}}$ is increased enough to drive $\tau_{\text{sr},\,\text{ann}} \ll T$, then $h_s$ only depends on the combination $R_{\text{sr},\,\text{ann}} \tau_{\text{sr},\,\text{ann}} = N_\text{sr}$ and becomes \textit{independent} of $\Gamma_{\text{sr},\,\text{ann}}$. Therefore the peak $h_s$, which occurs at large $M$ when $\tau_{\text{sr},\,\text{ann}} \ll T$, is independent of $\Gamma_{\text{sr},\,\text{ann}}$.

\subsection{Black Hole Inspiral}
\label{subsubsec:BH_merger}

In addition to superradiant annihilation discussed in Sec.~\ref{subsubsec:superradiance}, high-frequency GWs may also be produced from the inspiral of compact objects such as BHs, boson and fermion stars~\cite{Palenzuela:2007dm,Giudice:2016zpa,Palenzuela:2017kcg,Helfer:2018vtq}, gravitino stars~\cite{Narain:2006kx}, gravistars~\cite{Mazur:2004fk}, and DM blobs~\cite{Diamond:2021dth}. For simplicity we will examine the most straightforward scenario, the GW emission from the inspiral of two BHs with equal mass, $M$, in a perfectly circular orbit~\cite{maggiore}. To lose energy to GWs the orbital radius must decrease, drawing the BHs closer and thus increasing the frequency of the GWs produced. The minimum radius for which we can describe the inspiral as an adiabatically-changing circular orbit is, $R_\text{ISCO} = 12 G M$, where ISCO stands for ``Innermost Stable Circular Orbit" (ISCO). Before the BHs reach the ISCO the GW frequency is given by, $\omega\sim 10 \,{\rm meV}\, \left(10^{-9}M_\odot/M\right) \left(R_{\rm ISCO}/R\right)^{3/2}$. Requiring that $R > R_\text{ISCO}$ bounds $M \omega$ to be small. Therefore to reach a given ISCO frequency the BH mass must be less than $M_\text{max}^{\text{bhi}} \sim 10^{-8} \, M_\odot \, \left( \text{meV} /\omega_\text{ISCO} \right)$, where ``bhi" is shorthand for Black Hole Inspiral. Similar to the superradiant annihilation signal discussed in Sec.~\ref{subsubsec:superradiance}, generating high-frequency signals comes at the cost of smaller masses.

A crucial difference between the superradiant annihilation signal and BH inspiral is that while superradiant annihilation is monochromatic, the GW frequency from BH inspiral is changing. This signal is therefore a chirp signal according to the classification discussed previously. The timescale of frequency change is given by, $\tau_f^\text{bhi}~=~f / \dot{f}~\sim~10^{-18} \, \text{yr} \, \left( 10^{-8} \, M_\odot/M \right)^{5/3} \left( \text{meV} / \omega \right)^{8/3}$, which, while relatively fast compared to the observation time, is still large enough compared to the phonon lifetime (assuming $\omega_\nu / \gamma_\nu \sim 100$) to justify treating the signal as deterministic. Moreover, since $\tau_f^\text{bhi} \ll T$ the minimum frequency, $\omega_\text{min} = \omega(t = 0)$ will be much less than $\omega_\text{max} = \omega_\text{ISCO}$, allowing us to approximate the signal bandwidth as $0 \lesssim \omega \lesssim \omega_\text{ISCO}$. 

Within this bandwidth the GW amplitude is~\cite{maggiore},
\begin{align}
    h_0^\text{bhi} & = \frac{2}{ r M_\text{pl} } \left( \frac{M}{M_\text{pl}} \right)^\frac{5}{3} \left( \frac{\omega}{M_\text{pl}} \right)^\frac{2}{3} \sim 10^{-17} \, \left( \frac{\text{AU}}{r} \right) \left( \frac{M}{10^{-8} \, M_\odot} \right)^\frac{5}{3} \left( \frac{\omega}{\text{meV}} \right)^\frac{2}{3} \, ,
\end{align}
which, while seemingly large, is penalized in its overall signal strength ($h_s$ in Eq.~\eqref{eq:h_s_chirp}) by its small frequency coherence time, $\tau_f$. The signal strength is given by
\begin{align}
    h_s^\text{bhi} & = \sqrt{\frac{\tau_f^\text{bhi}}{T}} \, h_0^\text{bhi} \sim 10^{-26} \, \left( \frac{\text{AU}}{r} \right) \left( \frac{\text{yr}}{T} \right)^{\frac{1}{2}} \left( \frac{M}{10^{-8} \, M_\odot} \right)^\frac{5}{6} \left( \frac{\text{meV}}{\omega} \right)^\frac{2}{3} \, , \label{eq:h_s_bhm}
\end{align}
and shown in Fig.~\ref{fig:h_s_comparison} (labelled ``Black Hole Inspiral") as a function of BH mass for different choices of $\omega$ assuming $T = 1 \, \text{yr}$. 

\section{Conclusion}
\label{sec:conclusion}

Single-phonon excitations, with energies in the $\mathcal{O}(1 - 100) \, \text{meV}$ range, have been shown to be exceptionally sensitive to light DM~\cite{Schutz:2016tid,Knapen:2017ekk,Griffin:2018bjn,Trickle:2019nya,Cox:2019cod,Griffin:2019mvc,Trickle:2020oki,Mitridate:2020kly,Coskuner:2021qxo,Knapen:2021bwg,Mitridate:2023izi,Berlin:2023ppd}. Here we show that this sensitivity extends to high-frequency GWs in the $10^{12} \, \text{Hz} \lesssim f \lesssim 10^{14} \, \text{Hz}$ range. In Sec.~\ref{sec:theory_derivation} we derived the forces acting on a lattice of point masses due to an incoming GW, and then used this to derive the GW-single phonon interaction Hamiltonian given in Eq.~\eqref{eq:GW_phonon_interaction_Hamiltonian_0}. We then used this interaction to derive the absorption rate in non-polar, Sec.~\ref{subsubsec:non_polar_targets}, and polar, Sec.~\ref{subsubsec:polar_targets}, crystal targets. An important difference between signals generated by absorption of light DM and GWs is that the GW signal may not be monochromatic or coherent over the observation time. This introduces subtleties in characterizing the detector sensitivity relative to the signal strength. In Sec.~\ref{sec:results} we define the detector sensitivity, $h_\text{det}$, and signal strength, $h_s$, for the deterministic (as opposed to stochastic) signals of interest here. Most of the discussion readily generalizes to other direct detection experiments, making the definitions of $h_\text{det}$ and $h_s$ useful outside the context of just single-phonon excitations. We then computed the GW absorption rate into single-phonon excitations from first principles in nine targets: \ce{GaAs}, \ce{Al2O3}, \ce{SiO2}, \ce{Si}, \ce{Ge}, Diamond, \ce{NaI}, \ce{CsI}, \ce{CaWO4}, which have been well studied as targets in DM direct detection experiments. In Fig.~\ref{fig:monochromatic_detail_projections} we show the corresponding detector sensitivity for each individual target, and in Fig.~\ref{fig:monochromatic_overview_projections} we compare the detector sensitivities to other high-frequency GW experiments in the $10^6 \, \text{Hz} \lesssim f \lesssim 10^{14} \, \text{Hz}$ frequency range. The diversity of targets highlights how a combination of targets may be used to provide broadband coverage in GW frequency space.

In Sec.~\ref{subsec:mono_sources} we studied potential sources of these high-frequency GWs. GWs from the early universe are far too weak to be detectable with single-phonon detection, and an extraordinary amount of power must be emitted in GWs at $\mathcal{O}(\text{kpc})$ distances to be measurable. Therefore it seems any viable source must be within the local solar system, at $\mathcal{O}(\text{AU})$ distances. As examples of sources, we studied superradiant annihilation, Sec.~\ref{subsubsec:superradiance}, and BH inspiral, Sec.~\ref{subsubsec:BH_merger}, and illustrated their relative signal strength in Fig.~\ref{fig:h_s_comparison}. In addition to these specific examples, in Figs.~\ref{fig:monochromatic_detail_projections} and~\ref{fig:monochromatic_overview_projections} we show the signal strength corresponding to \textit{any} source emitting monochromatic, coherent, GWs with power $P_\text{GW}$. These serve as useful benchmarks since they provide both a physically motivated comparison between experiments, and a target for future studies of high-frequency GW sources.

While our focus has been on single-phonon excitations in crystal targets, there are other ways GWs may interact with phonons. For single-phonon excitations, the lowest accessible frequency is set by the energy of the lowest gapped mode. However gapless, acoustic phonon modes have even smaller energies. These can be utilized in multi-phonon processes in both crystal~\cite{Campbell-Deem:2019hdx}, and liquid~\cite{Knapen:2016cue,Baym:2020uos} targets where the incoming GW is kinematically matched to two phonons. It may also be interesting to utilize photon read out strategies as studied in Ref.~\cite{Marsh:2022fmo}. Furthermore, future research studying the details of different GW signals are important for differentiation from a DM induced signal. 

The introduction of external electromagnetic fields may also be advantageous. The same inverse Gertsenshtein effect which the axion DM experiments are utilizing may also be an avenue to excite phonons in crystal targets or optomechanical experiments~\cite{Murgui:2022zvy,Baker:2023kwz}. Indeed, if an incoming GW converts to an electromagnetic field, that electromagnetic field can then excite optical phonon modes in polar materials, analogous to axion absorption on phonons in magnetized media~\cite{Mitridate:2020kly,Berlin:2023ppd}. However in the crystal targets of interest here, the rate seems to be parameterically suppressed. The effective coupling parameter is $h_0 B_0$, where $B_0$ is the external magnetic field. When compared to the axion DM coupling, $g_{a \gamma \gamma} \sqrt{\rho_\text{DM}} B_0 / m_a$, this corresponds to $h_0 \sim g_{a\gamma\gamma}\sqrt{\rho_\text{DM}} / m_a \sim 10^{-22} \, g_{a\gamma\gamma} / \left( 10^{-12} \, \text{GeV}^{-1} \right)$ at $\omega \approx m_a = 10 \, \text{meV}$, leading to much worse detector sensitivity than shown here.

We have shown that DM direct detection experiments utilizing single-phonon excitations can be powerful probes of high-frequency GWs in the $\text{THz} \lesssim f \lesssim 100 \, \text{THz}$ frequency range. More broadly this suggests a connection between DM direct detection and high-frequency GW detection beyond axion experiments. This is important since the understanding of high-frequency GW sources is still emerging, and therefore multi-purpose DM-GW experiments present the best opportunity to explore the high frequency frontier. Furthermore, demonstrating sensitivity with multi-purpose DM-GW experiments motivates further research of GW sources which populate the high frequency landscape. 

\acknowledgements

We would like to thank Asher Berlin, Diego Blas, Valerie Domcke, Sebastian A. R. Ellis, Dan Hooper, Gordan Krnjaic, and Alex Millar for helpful discussions. 
The work of Y. Kahn and J. Sch\"utte-Engel was supported in part by DOE grant DE-SC0015655. J. Sch\"utte-Engel  acknowledges support from the RIKEN-Berkeley Center and from the National Science 
Foundation under cooperative agreement 2020275. TT's contribution to this work is supported by the Fermi Research Alliance, LLC under Contract No.\ DE-AC02-07CH11359 with the U.S.\ Department of Energy, Office of Science, Office of High Energy Physics.

\begin{figure}[ht!]
    \centering
    \includegraphics[width=\textwidth]{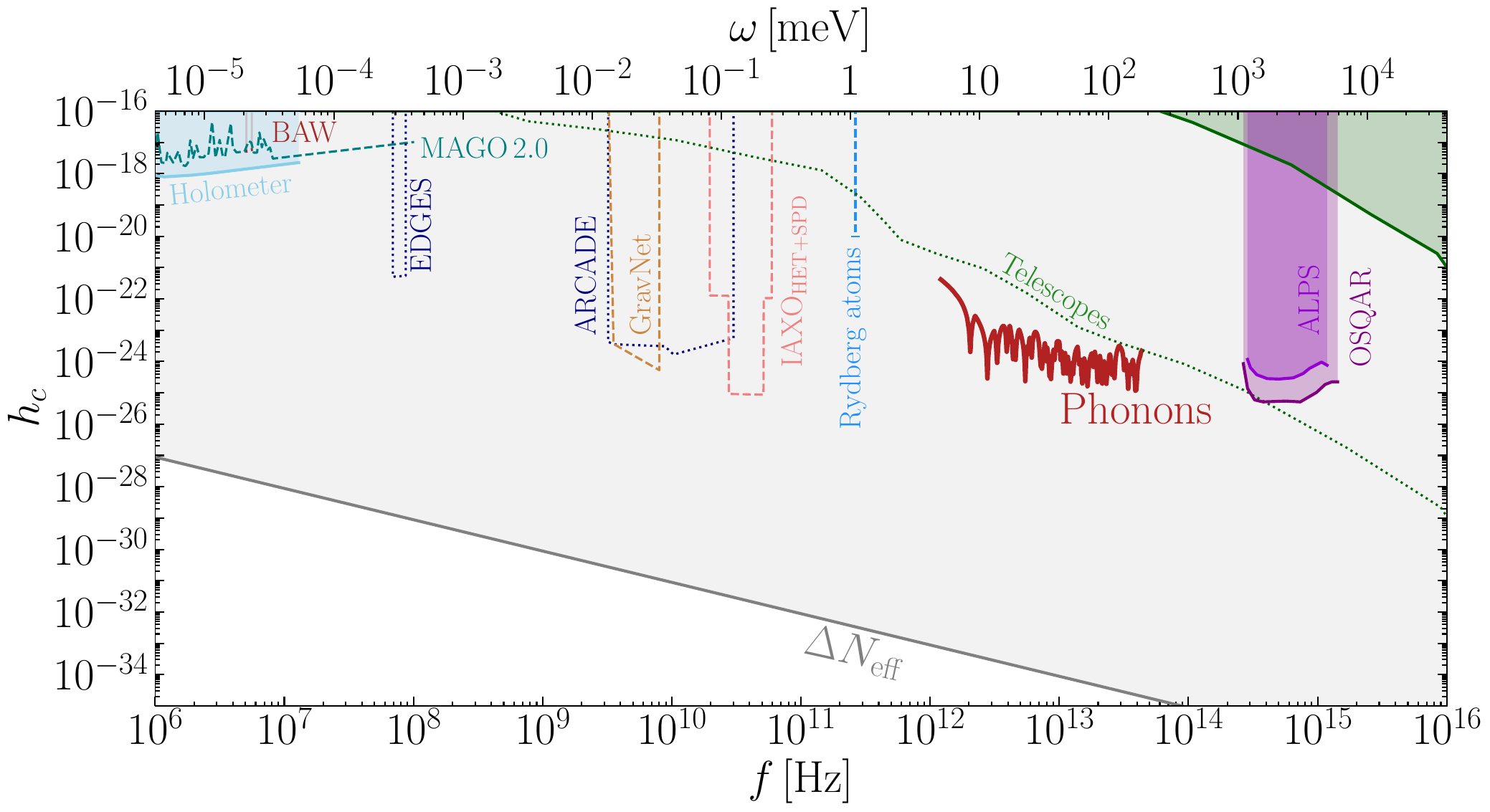}
    \caption{Overview of the constraints on the dimensionless strain, $h_c$, of stochastic GWs with frequency, $\text{MHz} \lesssim f \lesssim 10 \, \text{PHz}$. The solid red line labelled ``Phonons" is an outline of the detector sensitivities, $h_\text{det}$, shown in Fig.~\ref{fig:monochromatic_detail_projections}, assuming a $\text{kg} \cdot \text{yr}$ exposure and negligible backgrounds. Experimental constraints are shown in shaded regions, while projections are shown as dashed lines. Constraints from the Holometer experiment (light blue) are from Ref.~\cite{Holometer:2016qoh}, and those from bulk acoustic wave experiments (brown, labelled ``BAW") are from Ref.~\cite{Goryachev:2021zzn}. Constraints recast from the ``light shining through wall" experiments ALPS I~\cite{Ehret:2010mh} and OSQAR 2~\cite{OSQAR:2015qdv} (purple) are from Ref.~\cite{Ejlli:2019bqj}, and projections for future versions of IAXO utilizing HET and SPD detectors~\cite{Ringwald:2020ist} (light red, labelled ``IAXO$_{\text{HET}+\text{SPD}}$") are from Ref.~\cite{Ringwald:2020ist}. Projections for the MAGO 2.0 experiment are from Ref.~\cite{Berlin:2023grv} (teal). Projections utilizing systems of Rydberg atoms (blue) are from Ref.~\cite{Kanno:2023whr}. Projections using a global network of electromagnetic cavities (dark orange, labelled ``GravNet") are from Ref.~\cite{Schmieden:2023fzn}. Constraints from the EDGES~\cite{Bowman:2018yin} and ARCADE 2~\cite{Fixsen:2009xn} telescopes searching for the Gertsenshtein effect on CMB photons during the dark ages (dotted dark blue) are from Ref.~\cite{Domcke:2020yzq}, and assume optimistically large cosmic magnetic field strengths that saturate current CMB bounds. The dark green lines labelled ``Telescopes" correspond to constraints from many telescopes searching for photons converted by conservative (solid) and optimistic (dotted) values of the magnetic field in the intergalactic medium~\cite{Ito:2023nkq}. Additional telescopic constraints due to the Gertsenshtein effect in a variety of astrophysical magnetic fields, which generally lie within the range of the dotted and solid green lines, can be found in Refs.~\cite{Ito:2023fcr,Liu:2023mll,Ito:2023nkq}. The gray shaded region labelled ``$\Delta N_\text{eff}$" indicates the region bounded by a combination of BBN and CMB measurements on $\Delta N_\text{eff}$~\cite{Cyburt:2015mya,Aggarwal:2020olq}.}
    \label{fig:stochastic_overview_projections}
\end{figure}

\appendix

\section{Sensitivity to Stochastic Gravitational Waves}
\label{app:sto_constraints}

Having already discussed potential sources of stochastic high-frequency GWs in Sec.~\ref{subsec:mono_sources}, here we focus on computing the projected constraints, and compare them to other proposals and existing constraints in Fig.~\ref{fig:stochastic_overview_projections}. Stochastic GWs may be parameterized by either their energy density parameter, $\Omega_\text{GW}(f)$, or dimensionless, characteristic strain, $h_c$, related by Eq.~\eqref{eq:hc_strain_definition}. Assuming that $\Omega_\text{GW}(f)$ is independent of time, after multiplying Eq.~\eqref{eq:total_averaged_GW_abs_rate} by the detector mass and observation time, the number of phonons produced can be written in terms of the detector sensitivity, $h_\text{det}(f)$ from Eq.~\eqref{eq:h_det_definition}, as,
\begin{align}
    N_\text{ph} = 3 \times 4 \int \, \frac{h_c^2(f)}{h_\text{det}^2(f)} \, d \ln{f} \, . \label{eq:N_ph_stochastic}
\end{align}
Therefore, similar to the number of phonons produced from a chirp signal, Eq.~\eqref{eq:N_ph_chirp}, if $h_c(f)^2 \gtrsim h_\text{det}^2(f)$ over an e-fold in frequency, the stochastic signal will generate $\mathcal{O}(1)$ phonons. In Fig.~\ref{fig:stochastic_overview_projections} we compare an outline of the $h_\text{det}$ shown in Fig.~\ref{fig:monochromatic_detail_projections}, labelled ``Phonons", assuming a $\text{kg} \cdot \text{yr}$ exposure and negligible backgrounds, to a variety of other experiments. See the caption of Fig.~\ref{fig:stochastic_overview_projections} for specific references. 

In addition to the lines labelled ``EDGES/ARCADE", which are looking cosmic microwave background (CMB) photons perturbed by the Gertsenshtein effect during the dark ages~\cite{Domcke:2020yzq}, and the line labelled ``Telescopes", which corresponds to constraints from a collection of telescopes on photons produced by the Gertsenshtein effect in the intergalactic medium~\cite{Ito:2023nkq}, there are other constraints from telescopes. For example, Ref.~\cite{Ito:2023nkq} also considers photons converting in the Earth's magnetic field, and in the galactic magnetic field, Ref.~\cite{Ito:2023fcr} considers excess photons converted from magnetic fields within pulsars, and Ref.~\cite{Liu:2023mll} considers photon conversion in planetary magnetospheres. However the constraints not shown generally fall within the range of the dotted and solid green line (``Telescopes") shown in Fig.~\ref{fig:stochastic_overview_projections}, and are therefore omitted to avoid clutter. Furthermore, contrary to Fig.~\ref{fig:monochromatic_overview_projections}, constraints from other references are not rescaled, and taken directly from their respective source.

\bibliographystyle{utphys3}
\bibliography{biblio}

\end{document}